\documentclass[preprint2]{aastex}
\usepackage{lineno}

\catcode`\@=11
\newcommand{\gapprox}{\mathrel{\mathpalette\@versim>}}
\newcommand{\lapprox}{\mathrel{\mathpalette\@versim<}}
\newcommand{\propapprox}{\mathrel{\mathpalette\@versim\propto}}
\newcommand{\@versim}[2]
  {\lower3.1truept\vbox{\baselineskip0pt\lineskip0.5truept
\ialign{$\m@th#1\hfil##\hfil$\crcr#2\crcr\sim\crcr}}}
\catcode`\@=12

\shorttitle{MODELING THE GAMMA-RAY PULSAR POPULATION}
\shortauthors{PERERA ET AL.}

\begin{document}

\title{Modeling the non-recycled {\it Fermi} gamma-ray pulsar population}

\author{B.~B.~P.~Perera\altaffilmark{1}, M.~A.~McLaughlin\altaffilmark{1}, J.~M.~Cordes\altaffilmark{2}, M.~Kerr\altaffilmark{3}, T.~H.~Burnett\altaffilmark{4}, A.~K.~Harding\altaffilmark{5}}

\altaffiltext{1}{Department of Physics, West Virginia University,
Morgantown, WV 26506, USA.}
\altaffiltext{2}{Astronomy Department and NAIC, Cornell University, Ithaca, NY 14853, USA.}
\altaffiltext{3}{W. W. Hansen Experimental Physics Laboratory, Kavli Institute for Particle Astrophysics and Cosmology, Department of Physics and SLAC National Accelerator Laboratory, Stanford University, Stanford, CA 94305, USA.}
\altaffiltext{4}{Department of Physics, University of Washington, Seattle, WA 98195-1560, USA.}
\altaffiltext{5}{NASA Goddard Space Flight Center, Greenbelt, MD 20771, USA.}

\vskip 1 truein

\newpage

\begin{abstract}
We use {\it Fermi Gamma-ray Space Telescope} detections and upper limits on non-recycled pulsars obtained from the {\it Large Area Telescope} (LAT) to constrain how the gamma-ray luminosity $L_\gamma$ depends on the period $P$ and the period derivative $\dot{P}$. We use a Bayesian analysis to calculate a best-fit luminosity law, or dependence of $L_\gamma$ on $P$ and $\dot{P}$, including different methods for modeling the beaming factor. An outer gap (OG) magnetosphere geometry provides the best-fit model, which is $L_\gamma \propto P^{-a} \dot{P}^{b}$ where $a=1.36\pm0.03$ and $b=0.44\pm0.02$, similar to but not identical to the commonly assumed $L_\gamma \propto \sqrt{\dot{E}} \propto P^{-1.5} \dot{P}^{0.5}$. Given upper limits on gamma-ray fluxes of currently known radio pulsars and using the OG model, we find that about 92\% of the radio-detected pulsars have gamma-ray beams that intersect our line of sight. By modeling the misalignment of radio and gamma-ray beams of these pulsars, we find an average gamma-ray beaming solid angle of about $3.7\pi$ for the OG model, assuming a uniform beam. Using LAT-measured diffuse fluxes, we place a $2\sigma$ upper limit on the average braking index and a $2\sigma$ lower limit on the average surface magnetic field strength of the pulsar population of $3.8$ and $3.2 \times 10^{10}$~G, respectively. We then predict the number of non-recycled pulsars detectable by the LAT based on our population model. Using the two-year sensitivity, we find that the LAT is capable of detecting emission from about 380 non-recycled pulsars, including 150 currently identified radio pulsars. Using the expected five-year sensitivity, about 620 non-recycled pulsars are detectable, including about 220 currently identified radio pulsars. We note that these predictions significantly depend on our model assumptions.    
\end{abstract}


\section{Introduction}
\label{intro}

The instrument EGRET on the {\it Compton Gamma Ray Observatory} detected pulsed gamma rays from six energetic rotation powered pulsars and three possible candidates \citep[see][for a review]{tho04a}. With the launch of the {\it Fermi Gamma-ray Space Telescope} in 2008, the number of gamma-ray pulsar detections  has increased significantly, with 46 gamma-ray pulsar (GRP) detections reported in the First {\it Fermi} Large Area Telescope Catalog of Gamma-ray Pulsars \citep[][hereafter 1PC]{aaa+10a} and 83 GRP detections in the  Second {\it Fermi} LAT Source Catalog \citep{naa+12}.  
It is therefore timely to use LAT detections to constrain the basic physics of pulsar gamma-ray emission and the relationship between gamma-ray and radio emission and begin to understand the GRP population. We note that the Second {\it Fermi} LAT Catalog of Gamma-ray Pulsars\footnote{http://fermi.gsfc.nasa.gov/ssc/data/access/lat/2nd\_PSR\_catalog/} \citep[][, hereafter 2PC]{2pc}, which reports 117 GRP detections, became public only recently, so that we were not able to include the results in this analysis.

The main power source for electromagnetic radiation from pulsars is the loss of rotational energy or spin-down luminosity ($\dot{E} = I\Omega\dot{\Omega}$, where $I$ and $\Omega$ are the moment of inertia and the angular frequency of the pulsar, respectively). 
The rotational kinetic energy is dissipated through mechanisms such as nonthermal radiation and a relativistic wind. Accordingly, the gamma-ray luminosity must be less  than the spin-down luminosity (i.e. $L_\gamma < \dot{E}$) and, equivalently, the gamma-ray efficiency $\eta=L_\gamma/\dot{E} < 1$. For rotation-driven pulsars, we might expect the gamma-ray luminosity to scale as $L_\gamma \propto \dot{E}$ or $L_\gamma \propto \Delta V \propto \sqrt{\dot{E}}$, where $\Delta V$ is the voltage drop across the polar cap (PC) region \citep[see][]{rs75}. The luminosities of radio pulsars seem to scale roughly with $\sqrt{\dot{E}}$ \citep{acc02,fk06}.

The origins of the radio and gamma-ray emission of pulsars are not well understood. However, there are different types of geometrical models put forward to explain the emission from pulsar magnetospheres \citep{rs75,rom96a,dr03,mh04a}. Some of these models are in good agreement with the observed properties of some pulsars, but a realistic physical model has not yet been proposed \citep[see ][]{lst+12}. The radio emission may be generated within the PC region of the pulsar close to the NS surface \citep[i.e. PC models,][]{rs75}. The high-energy gamma-ray emission may be generated in the outer regions of the neutron star magnetosphere. The common outer magnetosphere models are ``outer gap'' (OG) \citep{chr86a,cr92,cr94,rom96a}, ``two pole caustic'' (TPC) \citep{dr03}, and ``slot gap'' (SG) models \citep{mh04a}. However, in special cases such as Crab pulsar (1PC) and B$1937+21$ \citep{gjv+12}, the peaks of radio and gamma-ray pulse profiles are in close alignment, implying that both the radio and the high-energy emission are generated within the outer magnetosphere \citep{cr77b,mh96}.

The main purpose of this work is to determine how the observed luminosities of GRPs depend on spin-down properties. \citet{mc00} -- hereafter MC00 -- used pulsar detection, upper limit, and diffuse background measurements from EGRET (energies $> 100$ MeV) in a Bayesian likelihood analysis to constrain the luminosity law for GRPs and model the GRP population. Their best-fit luminosity law was $L_\gamma \propto P^{-1} \dot{P}^{0.8}$ (see Table~2 for parameter uncertainties). In their paper, the law was written in terms of the surface dipole magnetic field $B_{12}$, where $B_{12}=\sqrt{10^{15}P\dot{P}}$~G. This law is similar to but inconsistent with $L_\gamma \propto \sqrt{\dot{E}}\propto P^{-1.5} \dot{P}^{0.5}$. \citet{mc03b} -- hereafter MC03 -- updated the model used in MC00 with distances from the electron density model of \citet{cl02}, resulting in a slightly different ($L_\gamma \propto P^{-1.1} \dot{P}^{0.6}$) best-fit law. Because of the large number of GRP detections reported with {\it Fermi}, we can improve the luminosity law and also constrain other properties of GRPs, such as the beaming fraction and the degree of misalignment of the radio and gamma-ray beams. We constrain the luminosity law only from fluxes of pulsar detections, instead of using both detections and upper limits as in MC00 and MC03. That is because the large number of upper limit measurements biases and dominates the likelihood analysis, resulting in lower model-estimated luminosities for detections.

In addition to the detected fluxes of pulsars, the luminosity law depends on pulsar distance and beaming solid angle ($\Omega_\gamma$), or the solid angle swept out by the gamma-ray beam. We are able to use the estimated distances of pulsars using various methods, albeit with large errors. 
We assume a flat distribution for pulsar distances within the errors resulting from all models.
Both MC00 and MC03 assumed a beaming solid angle of $2\pi$. In this work, in addition to using a constant beaming solid angle, we model the beaming solid angle individually for each pulsar according to its spin properties and observed pulse properties. We further investigate the beaming solid angle with  geometry estimated from radio polarization, if available, and also with different  geometrical emission models. However, there are few pulsars with geometry estimated from polarization, resulting in a small number of pulsars with known geometry \citep{ran93,mr11}.

We then use the constrained luminosity law in two analyses. First, we calculate gamma-ray upper limits for radio-detected pulsars and assume that their gamma-ray beam is out of our line of sight if the model-estimated flux from the luminosity law is much larger than the calculated upper limit. This idea is consistent with the suggestion of \citet{rkc+11} that sub-luminous gamma-ray pulsars have gamma-ray radiation that is beamed away from our line of sight. We then use this information to model the misalignment of the radio and gamma-ray beams and then estimate the average gamma-ray beaming solid angle. Secondly, we use our best-fit luminosity law in the GRP population model with measured {\it Fermi} diffuse fluxes to constrain some properties of the population such as braking index {\it n}, magnetic field {\it B}, and initial spin period {\it $P_0$}. In the population analysis of MC00, a single value for the magnetic field and the initial spin period of the population were used. However, as in MC03 and other studies \citep{fk06}, we use a log-normal distribution for the surface  magnetic field. We use a flat distribution for initial spin periods since the neutron star initial spin period distribution is not clearly understood \citep[for example, see][]{lbdh93}.

In Section~\ref{data}, we describe the {\it Fermi} LAT gamma-ray data that we use in our analysis and the upper limit calculation method. In Section~\ref{lum}, we explain the general form of the GRP luminosity law  and in Section~\ref{like} we discuss the maximum likelihood method that we use to constrain the best-fit parameters. Our results, using different methods for modeling the beaming solid angle, are discussed in Section~\ref{results}. In Section~\ref{beam}, we use gamma-ray upper limits of radio-detected pulsar to model the misalignment angle of the radio and gamma-ray beams. In Section~\ref{pop}, we discuss our model Galactic GRP population and the results of fitting the model-estimated fluxes to measured diffuse fluxes due to pulsars. Finally in Section~\ref{dis}, we discuss our results, including the GRP population detectable with {\it Fermi} LAT.

\section{Data}
\label{data}

\subsection{Pulsed gamma-ray detections from GRPs}
We use the pulsed gamma-ray detections from non-recycled pulsars reported in the 1PC and several other detections since then. The recycled millisecond pulsars (MSPs) have a different spatial distribution compared to that of young non-recycled pulsars. Further, their smaller magnetospheres may lead to different emission processes and luminosity laws. Thus, we exclude MSPs from our analysis and will present their results in a future work. All the detections we use in this analysis are given in Table~\ref{detect}.

\begin{table*}
\caption{
The gamma-ray detections used in the analysis. The second and third columns are timing-derived period and period derivative, respectively, for the 35 detections. The fourth column is the energy flux for $E>100$~MeV. The fifth column gives the distance estimate used, and the last column is the corresponding reference. Note that for most of the pulsars, we use the DM-derived distance from the NE2001 model \citep{cl02}.}
\begin{center}
\begin{tabular}{crrrrr}
\hline
\multicolumn{1}{c}{PSR} &
\multicolumn{1}{c}{P} &
\multicolumn{1}{c}{$\dot{P}$} & 
\multicolumn{1}{c}{Energy Flux (G)} &
\multicolumn{1}{c}{Distance} &
\multicolumn{1}{c}{Ref.} \\
\multicolumn{1}{c}{} &
\multicolumn{1}{c}{(ms)} &
\multicolumn{1}{c}{($10^{-15}$~s/s)} &
\multicolumn{1}{c}{($10^{-11}$~erg~cm$^{-2}$~s$^{-1}$)} &
\multicolumn{1}{c}{(kpc)} &
\multicolumn{1}{c}{} \\
\hline 
J0007$+$7303	& 316.0 & 361.0 & 38.20$\pm$1.30 & 1.40$\pm$0.30 & (1) \\
J0106$+$4855	&  83.2 &   0.4 &  1.93$\pm$0.18 & 3.09$\pm$0.93 & (2) \\
J0205$+$6449    &  65.7 & 194.0 &  6.64$\pm$0.65 & 4.50$\pm$1.35 & (3) \\
J0248$+$6021	& 217.0 &  55.1 &  3.07$\pm$0.70 & 2--9          & (1) \\
J0534$+$2200    &  33.1 & 423.0 &130.60$\pm$3.40 & 1.73$\pm$0.52 & (3) \\
J0631$+$1036    & 288.0 & 105.0 &  3.04$\pm$0.61 & 3.63$\pm$1.09 & (3) \\
J0633$+$1746    & 237.0 &  11.0 &338.10$\pm$3.50 & 0.250$^{+0.120}_{-0.062}$ & (4) \\
J0742$-$2822    & 167.0 &  16.8 &  1.82$\pm$0.42 & 2--7          & (5) \\
J0835$-$4510	&  89.3 & 124.0 &879.40$\pm$5.40 & 0.287$^{+0.019}_{-0.017}$ & (6) \\
J1028$-$5819    &  91.4 &  16.1 & 17.70$\pm$1.40 & 2.33$\pm$0.70 & (3) \\
J1048$-$5832    & 124.0 &  96.3 & 17.20$\pm$1.30 & 2.71$\pm$0.81 & (3) \\
J1057$-$5226    & 197.0 &   5.8 & 27.20$\pm$0.98 & 0.72$\pm$0.20 & (3) \\
J1119$-$6127    & 408.7 &4027.8 &  7.10$\pm$0.50 & 8.40$\pm$0.40 & (7) \\
J1124$-$5916    & 135.0 & 747.0 &  3.79$\pm$0.70 & 5.70$\pm$1.71 & (3) \\
J1357$-$6429    & 166.2 & 357.2 &  3.39$\pm$0.33 & 2.50$\pm$0.75 & (3) \\
J1418$-$6058    & 111.0 & 170.0 & 23.50$\pm$3.80 & 2--5          & (1) \\
J1420$-$6048    &  68.2 &  83.2 & 15.80$\pm$3.50 & 5.60$\pm$1.70 & (3) \\
J1509$-$5850    &  88.9 &   9.2 &  9.70$\pm$1.20 & 2.60$\pm$0.80 & (3) \\
J1709$-$4429    & 102.0 &  93.0 &124.00$\pm$2.60 & 2.30$\pm$0.69 & (3) \\
J1718$-$3825    &  74.7 &  13.2 &  6.70$\pm$1.90 & 3.82$\pm$1.15 & (3) \\
J1732$-$3131    & 196.5 &  28.0 & 24.20$\pm$1.40 & 0.61$\pm$0.18 & (3) \\
J1741$-$2054    & 414.0 &  16.9 & 12.80$\pm$0.80 & 0.38$\pm$0.11 & (3) \\
J1747$-$2958    &  98.8 &  61.3 & 13.10$\pm$1.70 & 2.00$\pm$0.60 & (3) \\
J1809$-$2332    & 147.0 &  34.4 & 41.30$\pm$1.60 & 1.70$\pm$1.00 & (1) \\
J1833$-$1034    &  61.9 & 202.0 & 10.10$\pm$1.40 & 3.30$\pm$0.99 & (3) \\
J1836$+$5925    & 173.3 &   1.5 & 59.90$\pm$1.30 & 0.50$\pm$0.15 & (3) \\
J1907$+$0602    & 106.6 &  86.7 & 25.40$\pm$0.60 & 3.21$\pm$0.96 & (3) \\
J1952$+$3252    &  39.5 &   5.8 & 13.40$\pm$0.90 & 3.14$\pm$0.94 & (3) \\
J2021$+$3651    & 104.0 &  95.6 & 47.00$\pm$1.80 & 2--4          & (8) \\
J2021$+$4026    & 256.0 &  54.8 & 97.60$\pm$2.00 & 1.50$\pm$0.45 & (3) \\
J2030$+$3641    & 200.1 &   6.5 &  3.14$\pm$0.33 & 2--4          & (9) \\
J2032$+$4127    & 143.0 &  19.6 & 11.10$\pm$1.40 & 3.60$\pm$1.08 & (3) \\
J2043$+$2740    &  96.1 &   1.3 &  1.55$\pm$0.32 & 1.80$\pm$0.54 & (3) \\
J2229$+$6114    &  51.6 &  78.3 & 22.00$\pm$1.00 & 7.50$\pm$2.25 & (3) \\
J2240$+$5832    & 139.9 &  15.2 &  1.08$\pm$0.32 &10.18$\pm$3.05 & (3) \\
\hline \end{tabular}
\tablecomments{
References: (1) 1PC -- \citet{aaa+10a}; (2) \citet{pga+12}; (3) NE2001 -- \citet{cl02}; (4) \citet{fwa+07}; (5) \citet{waa+10}; (6) \citet{dlrm03}; (7) \citet{pkd+11}; (8) \citet{aaa+09b}; (9) \citet{ckr+12} 
}
\end{center}
\label{detect}
\end{table*}

The distance to each pulsar is required by our analysis. From the 38 non-recycled detections in the 1PC, 22  are detected at radio frequencies and we therefore estimate their distances using dispersion measures (DMs) coupled with the NE2001 electron density  model \citep[][]{cl02}, with the following exceptions. Note that we ignored PSR J$0659$$+$$1414$ in the analysis due to its extremely low gamma-ray efficiency \citep{waa+10,twc+11}, which resulted in an extreme outlier in our analysis. We use the parallax-estimated distance for PSR J$0835$$-$$4510$ \citep{dlrm03} and the 1PC reported distance for PSR J$0248$$+$$6021$ due to its unreliable DM-derived distance. For PSR J$2021$$+$$3651$, the DM-derived distance is 12~kpc. As argued in \citet{aaa+09b}, such a large distance implies an unphysically  high  gamma-ray efficiency  for some beaming models. Therefore, we use the distance that they derived, 2--4~kpc, from the pulsar wind nebula properties of the X-ray observations. Furthermore, the DM-derived distance for PSR J$0742$$-$$2822$ is 2.1~kpc. However, due to excess density from the Gum Nebula, we use the kinematic distance to the pulsar, 2--7~kpc, as discussed in \citet{waa+10}. In addition to these radio-loud pulsars, we adopt the distances reported in the 1PC (see Table~5 therein) for radio-quiet PSRs J$0633$$+$$1746$, J$0007$$+$$7303$, J$1809$$-$$2332$, J$2021$$+$$4026$, and J$1418$$-$$6058$. Note that PSR J$0633$$+$$1746$ has a parallax-estimated distance \citep{fwa+07}. Therefore, 27 (i.e., 22 radio-loud and 5 radio-quiet pulsars) out of 38 non-recycled pulsar detections reported in the 1PC have distance estimates.

Several new pulsed gamma-ray detections  with distance estimates have been reported since the 1PC was published. These include PSRs J$0106$$+$$4855$ \citep{pga+12}, J$1119$$-$$6127$ \citep{pkd+11}, J$1357$$-$$6429$ \citep{lzg+11}, J$2030$$+$$3641$ \citep{ckr+12}, and J$2240$$+$$5832$ \citep{tpc+11}. Further, we make use of updated distance estimates for three pulsars; PSRs J$1732$$-$$3131$, J$1907$$+$$0602$, and J$1836$$+$$5925$ \citep{aaa+10c,rkp+11}. The DM-derived distance to PSR J$1119$$-$$6127$ is $\sim$17~kpc, which places the pulsar beyond the Sagittarius arm. 
Therefore, we use the distance of $8.4$~kpc derived from HI absorption towards the supernova remnant \citep{cmc+04}. PSR J$2030$$+$$3641$ has a DM-derived distance of 8~kpc that is beyond the Cygnus region, which is known to have excess ionized gas that contributes to the DM and thus perturbs the NE2001 distance estimate. We use the most likely distance range of 2--4~kpc as mentioned in \citet{ckr+12}.
Thus, in total, there are 35 GRP detections in our sample; 27 from the 1PC and eight additional detections. For all DM-derived distances, we assumed an uncertainty of 30\%.

The pulsed gamma-ray energy fluxes for the above 35 GRPs are taken from the 1PC and the above mentioned papers. All these fluxes were obtained by fitting an exponentially-cutoff power-law model to the pulsar spectra with the energy range of 100~MeV $<$ E $<$ 100~GeV. The measured energy fluxes of these 35 pulsars range from $(1.1-879)\times 10^{-11}$~erg s$^{-1}$ cm$^{-2}$.

\subsection{Gamma-ray flux upper limits}
\label{upperlimit}

In order to model the misalignment of radio and gamma-ray beams of radio-detected pulsars, we compute their gamma-ray upper limits using three years of Pass 7 LAT data, accepting events with zenith angles $<100\degr$ and with energies between 100~MeV and 1~TeV. The data are binned in energy (logarithmically, four per decade) and spatially using HEALPix\footnote{http://healpix.jpl.nasa.gov} \citep{ghb+05}, and likelihood analysis is performed with {\it pointlike} \citep{ker10}, which has been shown to yield results consistent with the publicly-released LAT Science Tools analysis package. The gamma-ray background is modeled using the same diffuse models\footnote{http://fermi.gsfc.nasa.gov/ssc/data/access/lat/BackgroundModels.html} as the second {\it Fermi} source catalog \citep[2FGL; ][]{naa+12} and an internal list of point sources based on the three-year data set. To compute the upper limit, we tessellate the sky into HEALPixels with $nside=512$ (resolution $\sim$1.6 arcmin) and test for the presence of a point source in each of these pixels. We assume a representative pulsar spectral shape, a power law with index 1.8 and an exponential cutoff with cutoff energy 2~GeV.  We vary the spectral normalization until the likelihood decreases to the 95\% confidence level and thus determine the flux upper limit at the given position. Following this procedure, we calculate upper limits for 1496 non-recycled radio pulsars given in the ATNF pulsar catalog\footnote{http://www.atnf.csiro.au/people/pulsar/psrcat/} and include them in our analysis. We note that these upper limits are not entirely compatible with the point source fluxes reported in 1FGL \citep{aaa+10d}, which use an earlier version (Pass 6) of the LAT data and from which the luminosity law parameters are derived (see 2FGL for a discussion of fluxes). However, these are compatible with the data of 2PC. Because errors are dominated by uncertainty in distance, we do not expect these small differences to substantially affect results derived with the upper limits.

\subsection{Gamma-ray diffuse flux measurements}

We use the LAT-team generated model of diffuse fluxes to study the Galactic GRP population assuming that a fraction of diffuse flux is due to pulsars. The model is based on diffuse-class events recorded within the first eleven months of Pass 6 LAT data\footnote{http://fermi.gsfc.nasa.gov/ssc/data/access/lat/BackgroundModels.html} (Note that we used Pass 6 data because Pass 7 data were not available when this diffuse analysis was done).
The file $gll_{-}iem_{-}v02.fit$ contains the Galactic diffuse intensities as a function of Galactic latitude, longitude, and energy after subtracting the contribution of point sources. There are 30 logarithmically spaced energy bins between 50~MeV and 100~GeV in the data file \citep{aaa+12}. We use the given Galactic diffuse emission differential intensity (photons s$^{-1}$ sr$^{-1}$ cm$^{-2}$ Mev$^{-1}$) to calculate the diffuse fluxes (erg s$^{-1}$ sr$^{-1}$ kpc$^{-2}$). For our population study, we calculate the diffuse fluxes in four directions along the Galactic plane: $1.8\times10^{37}$, $1.3\times10^{37}$, $8.0\times10^{36}$, and $3.5\times10^{36}$~erg s$^{-1}$ sr$^{-1}$ kpc$^{-2}$ along $l = 0\degr$, $20\degr$, $40\degr$, and $60\degr$, respectively. Because the diffuse background models and pulsar spatial distribution are associated with large uncertainties, this coarse binning is sufficient for our work. We assume 25\% uncertainties on these fluxes, keeping in mind that the uncertainties associated with our assumption of the pulsar contribution to the diffuse flux ($\leq$10\%, see Section~\ref{pop}) are much greater.

\section{Luminosity law}
\label{lum}

The spin-down luminosity is $\dot{E}=4\pi^2 I P^{-3}\dot{P}$, where $I=10^{45}$~g~cm$^2$ is the typical value used for the moment of inertia, assuming a uniform sphere with a radius of 10~km and a mass of 1.4~M$_\odot$. Therefore, the gamma-ray luminosity can be written as a function of $P$ and $\dot{P}$. As discussed in Section~\ref{intro} and given in Table~2, pulsar luminosities scale roughly with $\sqrt{\dot{E}}$. However, previous studies such as MC00, MC03, and \citet{acc02} showed that the luminosity does not exactly follow the typical forms (see Table~2). To allow flexibility in our model and to be consistent with previous results, we parametrize the gamma-ray luminosity as

\begin{equation}
\label{luminosity}
L_\gamma = c P^{-a}\dot{P}_{15}^{b}~{\rm erg~s^{-1}},
\end{equation}

\noindent
where $a$, $b$, $c$ are constants and $\dot{P}_{15}=10^{15}\dot{P}$.

For a pulsar energy flux $G$, we write the luminosity as $L_{\rm \gamma} = 4\pi f_{\rm \Omega} D^2 G$, where $f_\Omega$ is the beaming factor and $D$ is the distance to the pulsar. The beaming factor is a geometrical term that is used to convert the energy flux to luminosity \citep{wrw+09}. In earlier work, $f_\Omega$ was taken to be $1/4\pi$ for a narrow gamma-ray emission beam of 1~sr \citep{tab+94}, or $f_\Omega=0.5$ for MSPs \citep{fab+95}. However, more recent theoretical work based on outer magnetosphere models shows that the gamma-ray emission beam sweeps nearly the entire celestial sphere, resulting in $f_\Omega \sim 1$ \citep{wrw+09}. On the other hand, \citet{twc+11a} argued that most of the emission in the OG model is within $90\degr \pm 35\degr$ of the rotation axis, indicating $f_\Omega \sim 35\degr/90\degr \sim 0.4$. In contrast, most radio pulsars have narrow PC radio emission beams with a best-fit half-opening angle of $\rho_{\rm r} = 5\fdg8 P^{-0.5}$ \citep{ran93}, where $P$ is in seconds, implying $f_\Omega \sim 0.1$ for an orthogonal rotator with spin period of 1 second.

\section{Luminosity law likelihood analysis}
\label{like}

In this section we describe how we use detections to calculate the best-fit luminosity law. First we calculate the luminosity of a detected pulsar $L_{\rm \gamma} (\Theta)$ from  Equation~(\ref{luminosity}) based on $P$ and $\dot{P}$ for a given model parameter combination $a$, $b$, and $c$ denoted as $\Theta$. Then we calculate the model-estimated energy flux $G_{\rm mod}(\Theta,D) = L_{\rm \gamma} (\Theta)/4\pi f_\Omega D^2$ for the estimated pulsar distance. The beaming factor $f_\Omega$ is modeled through different methods (see Section~\ref{results}). Then we calculate the likelihood of this model-estimated flux based on the pulsar's measured gamma-ray flux ($G$). We write the individual likelihood ${\cal L}_{\rm det,i}(\Theta)$ for a given model $\Theta$ for a given GRP $i$ as

\begin{equation}
\label{L_det_i}
{\cal L}_{\rm det,i}(\Theta) = \int_{D_{\rm l}}^{D_{\rm u}} dD f(D)g(\Theta,D)
\end{equation}

\noindent
where $g(\Theta,D) = (2\pi\sigma^2)^{-1/2} \exp (-(G_{\rm mod}(\Theta,D) - G)/2\sigma^2)$ and $f(D) = 1/(D_{\rm u}-D_{\rm l})$. The quantity $\sigma$ is the error on the measured energy flux and $D_{\rm l}$ and $D_{\rm u}$ are the lower and upper estimates on the distance.
Then we calculate the total likelihood from all GRPs for a given model $\Theta$ as

\begin{equation}  
{\cal L}_{\rm det}(\Theta) = \prod_{{\rm i=1}}^{N_{\rm det}} {\cal L}_{\rm det,i}(\Theta),
\end{equation}

\noindent
where $N_{\rm det}$ is the total number of detections.

We assume that the maximum possible luminosity of a pulsar is some fraction of the spin-down luminosity, i.e. $L_{\rm max} = \epsilon_\gamma \dot{E}$. We assume  $\epsilon_\gamma = 1$, in contrast to MC00's assumption of $\epsilon_\gamma = 0.5$. This is consistent with the highest realistic $\epsilon_\gamma$ for a pulsar detection assuming $f_\Omega = 1$ as in 1PC ($\epsilon_\gamma \approx 0.8$ for PSR J$0633+1746$). Note that as reported in the same study, PSR J$2021+4026$ has $\epsilon_\gamma \approx 2.2$ with $f_\Omega=1$, indicating that $f_\Omega$ for this pulsar is less than unity or the distance estimate is incorrect. If the model-estimated luminosity is greater than $\epsilon_\gamma \dot{E}$, we assign a likelihood of zero to that particular parameter combination for that pulsar. In contrast, if the model predicted luminosity was greater than $\epsilon_\gamma \dot{E}$, MC00 set the maximum luminosity equal to $\epsilon_\gamma \dot{E}$. Our modified method ensures that unrealistic models do not contribute to the likelihood.

We use a grid search  to determine the best-fit values for the three model parameters. For each combination of parameters, we calculate the likelihood for detections and then determine the marginalized probability density functions (PDFs) as

\begin{equation}
f(\Theta_{\rm j}) = \frac{\int_{\rm i \ne j} d\Theta_{\rm i} {\cal L}_{\rm det}(\Theta_{\rm i})}{\int d\Theta_{\rm i} {\cal L}_{\rm det}(\Theta_{\rm i})},
\label{pdfs}
\end{equation}

\noindent
where the numerator gives the total likelihood for any given parameter $\Theta_j$ across  each grid cell. The peak of this PDF gives the best-fit value of the parameter and then the error can be calculated for a desired confidence level.

\section{Luminosity law analysis and results}
\label{results}

We model the beaming factor using several different methods. MC00 assumed that the beaming solid angle $\Omega_\gamma$ of the gamma-ray emission is $2\pi$, or the beaming factor $f_\Omega = 0.5$,   meaning the gamma-ray emission covers half of the sky. If the emission is uniform across the instantaneous gamma-ray beam (i.e. the solid angle of the gamma-ray beam itself), then the maximum possible $f_\Omega$ would be unity, depending on the magnetic inclination and the angular radius of the beam. Note that the instantaneous gamma-ray beam solid angle ($\Omega_{\rm \gamma i}$) is smaller than the beaming solid angle $\Omega_\gamma$. However, if the emission is not uniform across the instantaneous gamma-ray beam and the true average flux across the beam itself is greater than what we have measured, then $f_\Omega$ can have values greater than unity. \citet{wrw+09} showed that for a given outer magnetosphere emission model, $f_\Omega$ can have a wide range of values depending on the geometry of the pulsar and the characteristic gap width.

As the first step, we model the beaming factor as a constant (Section~\ref{constant}), keeping in mind that it is likely  dependent on the emission geometry and the other pulsar properties. Therefore, in later sections, we use more sophisticated  models for $f_\Omega$.
In all these methods, we set a maximum value of $f_\Omega=1.5$ in order to be consistent with \citet{wrw+09}.

\subsection{Constant beaming factor}
\label{constant}

First, we use $f_\Omega=0.5$, which is then a similar analysis to MC00 and MC03, but using a larger sample of pulsar detections (35 compared to 7). By searching the entire parameter space for $a$, $b$, and $c$, we calculate the marginalized PDFs and $95\%$ confidence intervals to find $a = 1.43^{+0.03}_{-0.04}$, $b = 0.40\pm0.02$, and $\log c = 32.53\pm0.02$. Comparisons with previous and expected forms are given in Table~2. We list reduced chi-squared values for all fits in this table. These are all significantly greater than unity, indicating that none of our simple models are perfect fits to the data and that uncertainties in distance and beaming models dominate. Therefore, the uncertainties of our best-fit parameters are likely underestimated. The new luminosity law is substantially different from that of MC00. The new parameters have smaller errors compared to the previous estimates, likely due to the larger number of flux measurements. The last column of the table gives the number of severe non-detections for each model (see Section~\ref{beam} for more discussion). A ``severe non-detection" is defined as when the expected energy flux ($\hat{G}$) for an upper-limit pulsar from the best-fit luminosity law is greater than its measured upper limit flux ($G_{\rm up}$) by more than the 2$\sigma$ error (i.e. $\hat{G} > G_{\rm up} + 2\sigma$). In order to account for the distance uncertainty, we define the expected flux as $\hat{G} = L_{\rm \gamma}/4\pi f_{\rm \Omega} D_{\rm u}^2$, where $D_{\rm u}$ is the distance upper limit.

\begin{table*}
\begin{center}
\caption{
Constrained gamma-ray luminosity laws: $L_\gamma = c P^{-a} \dot{P}_{15}^{b}$. 
The fifth column is the log of the $\chi^2_{\rm red}$ value of the best-fit solution. The last column shows the number of severe non-detections out of 1496 pulsars for each model from the beaming analysis of the upper-limit pulsars.}
\begin{tabular}{lccccl}
\hline 
\multicolumn{1}{c}{} &
\multicolumn{1}{c}{$a$} &
\multicolumn{1}{c}{$b$} & 
\multicolumn{1}{c}{log$(c)$} &
\multicolumn{1}{c}{log($\chi^2_{\rm red}$)} &
\multicolumn{1}{l}{$N_{\rm \bar{\gamma}}$} \\
\hline
$L_\gamma \propto \dot{E}$ & 3 & 1 & - & - & - \\
$L_\gamma \propto \sqrt{\dot{E}}$ & 1.5 & 0.5 & - & - & - \\
\citet{mc00} & 1.0(1) & 0.8(2) & 32.0(2) & - & - \\
\citet{mc03b} & 1.1 & 0.6 & 32.4 & - & - \\
\citet{acc02} & 1.3(3) & 0.4(1) & 29.3(1)  & - & - \\
\hline
This work \\
\hline
Schematic beam models: &        \\
$f_\Omega=0.5$ & 1.43(4) & 0.40(2) & 32.53(2) & 2.94 & 114 \\
$f_\Omega=1$ & 1.43(4) & 0.40(3) & 32.83(2) & 2.93 & 114 \\
$\Omega_\gamma = \lambda P^{-\nu}$ & 0(1) & 0.5(1) & 33.3(3) & 5.1 &  94 \\
$\Omega_\gamma = \lambda (\Delta\phi) ^{\nu}$ & 2.0(2) & 0.4(2) & 32.0(6) & 4.54 & 41$^\dagger$ \\
Emission gap  models:&	\\
TPC & 1.45(4) & 0.41(2) & 32.81(3) & 2.93 & 162 \\
OG & 1.36(3) & 0.44(2) & 32.82(3) & 2.91 & 117 \\
PC & 1.11(4) & 0.38(2) & 32.74(1) & 3.49 & 162 \\
\hline 
\end{tabular}
\tablecomments{The number in parentheses is the 2$\sigma$ uncertainty in the last quoted digit. \\
$^\dagger$ Assuming a beaming factor of unity for pulsars  with flux upper-limits.\\ 
}
\end{center}
\label{full_result}
\end{table*}

\begin{figure*}
\epsscale{2.0}
\plotone{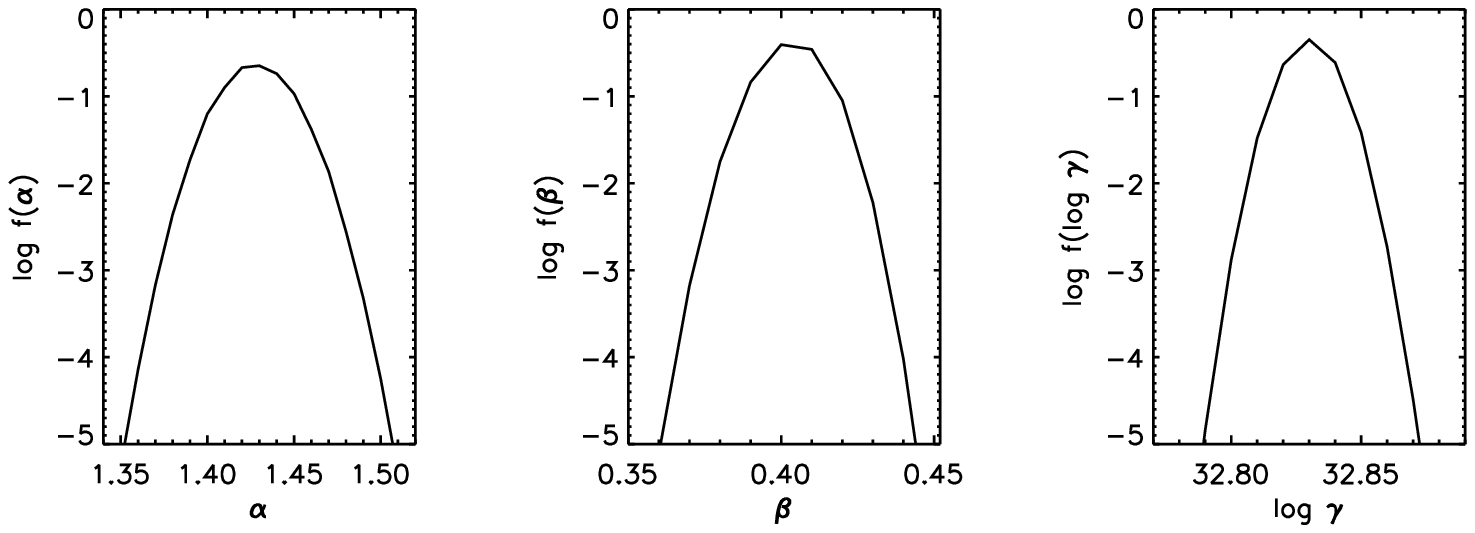}
\caption{
Marginalized PDFs for model parameters of the luminosity law when $f_\Omega=1$. The best-fit parameters at $95\%$ confidence level are $a = 1.43^{+0.03}_{-0.04}$, $b = 0.40^{+0.03}_{-0.02}$, and $\log c = 32.83\pm0.02$.
\label{pdf_4pi}}
\end{figure*}

\begin{figure*}
\epsscale{1.60}
\plotone{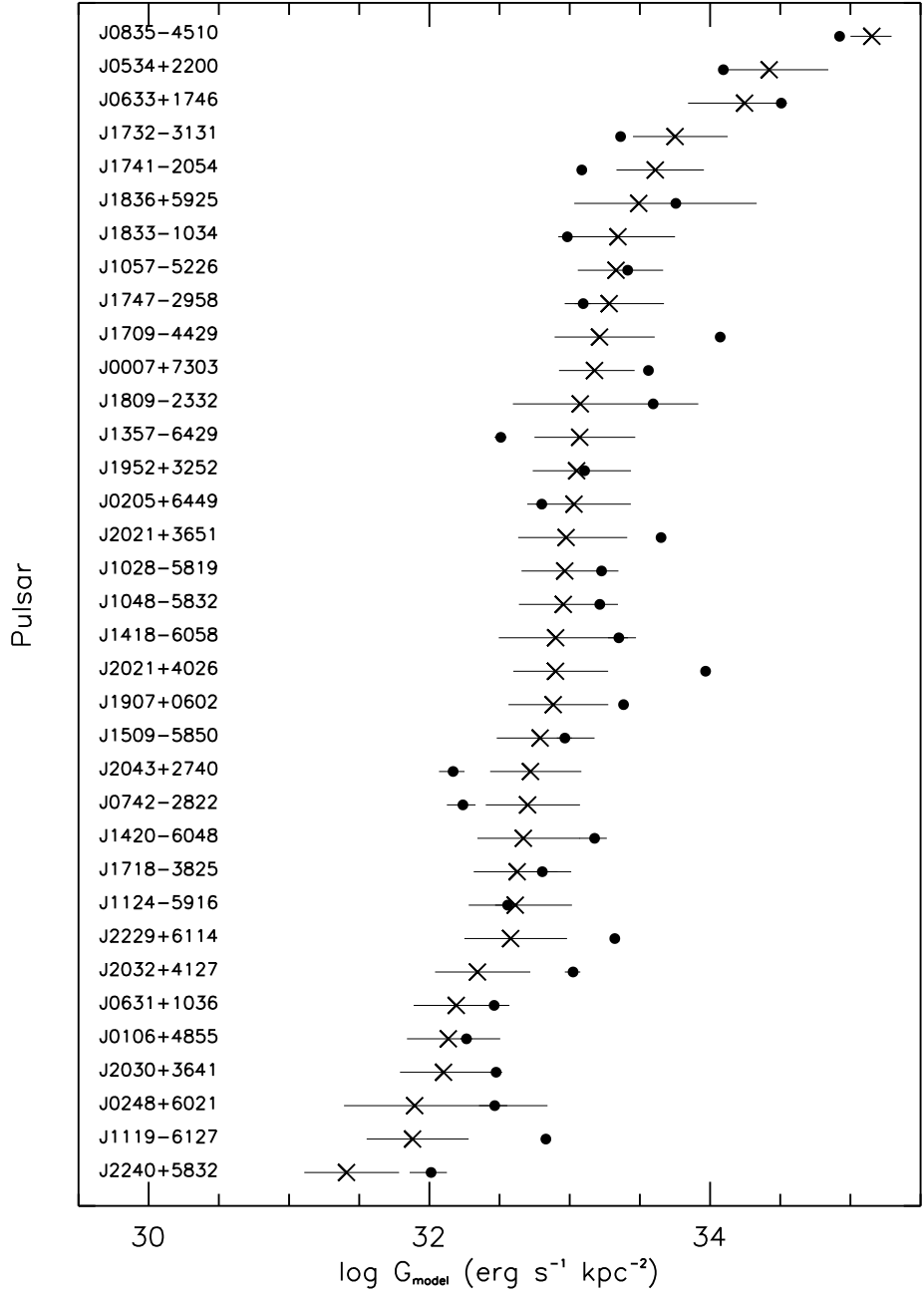}
\caption{
Model-estimated energy fluxes of 35 LAT-detected canonical pulsars with estimated distances for the case of a constant beaming factor, $f_\Omega=1$. The luminosity model that was used to calculate the model-estimated fluxes is $L_\gamma = 10^{32.83} P^{-1.43} \dot{P}^{0.40}$ (see Table~2). The model-estimated fluxes are marked by crosses and sorted in decreasing order. Dots represent the measured fluxes from the LAT. The errors of the model-estimated fluxes are determined through 2$\sigma$ errors of the three parameters $a$, $b$, and $c$ and the uncertainty of the distance. The errors of measured energy fluxes are much smaller than of the model-estimated energy fluxes, as the latter incorporates distance uncertainties. The logarithmic value of the $\chi^2_{red}$ is 2.93. Some of the outliers may be due to poorly estimated distance (see text).
\label{plot_flux}}
\end{figure*}

\begin{figure*}
\epsscale{1.70}
\plotone{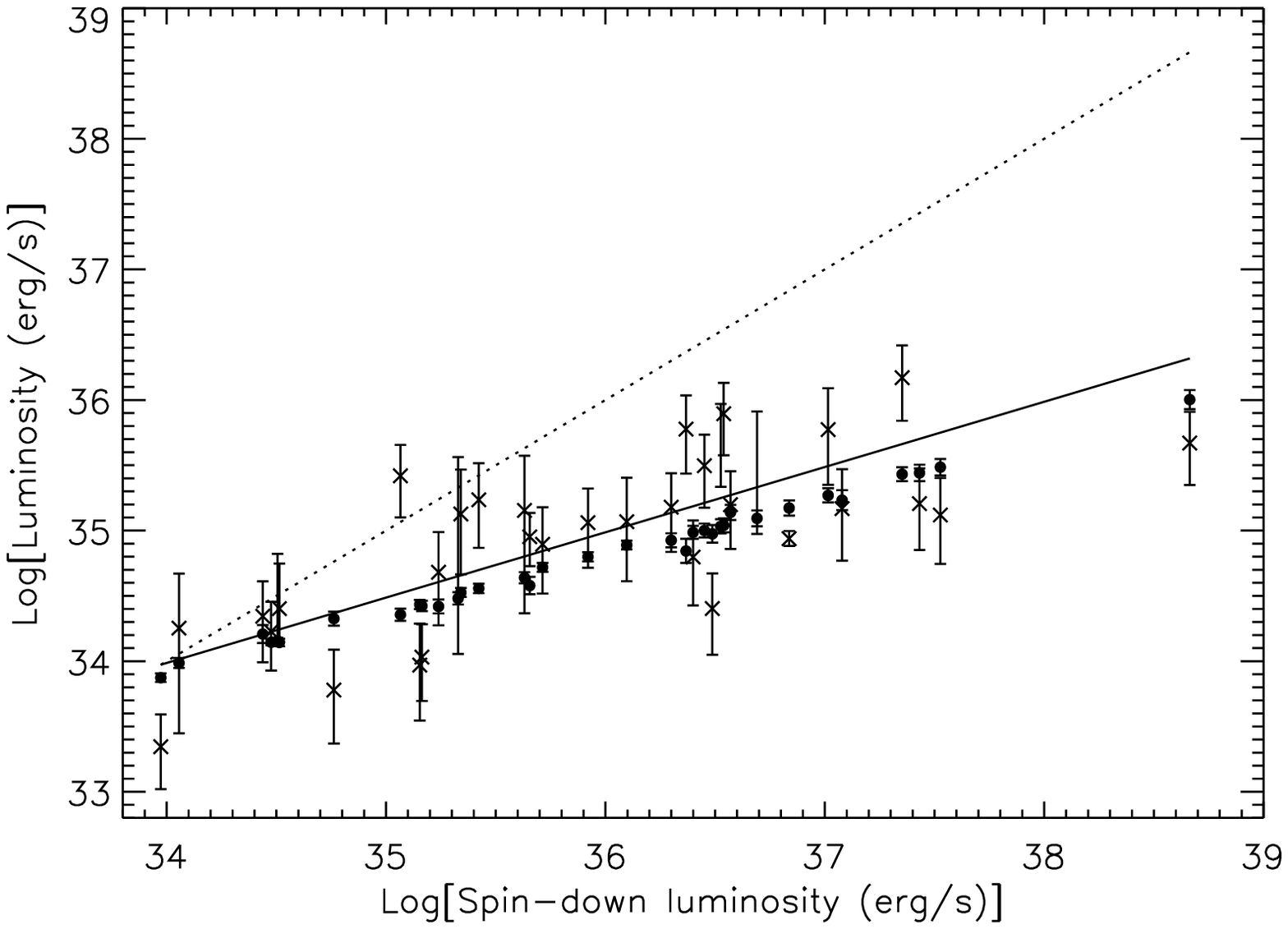}
\caption{
Model-estimated luminosity vs. spin-down luminosity ({\it Circles}) of the 35 LAT-detected pulsars for the case of a constant beaming factor, $f_\Omega=1$ (see Table~2 for the best-fit luminosity law). Note that the estimated errors are small due small errors on the best-fit luminosity parameters. The dotted line shows $L_\gamma = \dot{E}$. The solid line shows the form of $L_\gamma \propto \sqrt{\dot{E}}$. It is clearly seen that the model-estimated luminosities of these detections closely follow the latter form. For comparison, the computed luminosities for these pulsars based on the measured fluxes and estimated distances for $f_\Omega=1$ are shown ({\it Crosses}). The error of these luminosities are estimated from the uncertainties of distances and measured fluxes.
\label{lum_edot}}
\end{figure*}

We then examined the effect of using other constant values of the beaming factor. Most of the recent studies assume that $f_\Omega$ is unity \citep[1PC;][]{tpc+11,pga+12}. With this assumption, the marginalized PDFs of the three parameters are shown in Figure~\ref{pdf_4pi}. With the 95\% confidence interval, our best-fit model parameters are $a = 1.43^{+0.03}_{-0.04}$, $b = 0.40^{+0.03}_{-0.02}$, and $\log c = 32.83\pm0.02$ (see Table~2). Note that when the beaming factor is a constant, the luminosity scale ($c$) changes slightly while the model parameters ($a$ and $b$) remain nearly constant. Figure~\ref{plot_flux} shows the model-estimated and measured fluxes of these 35 GRPs. The errors of the model-estimated fluxes are calculated from the 2$\sigma$ errors of the three parameters and the errors on the distances. Figure~\ref{lum_edot} shows how the best-fit luminosities of these detections vary with their spin-down luminosities, with a luminosity law similar to $L_\gamma \propto \sqrt{\dot{E}}$. We assume this model as our reference model and use the particular luminosity law in the population analysis.

Note that the uncertainty in the model-estimated flux for J$0835$$-$$4510$ (Vela) is very small due to the well-constrained distance from parallax  \citep{dlrm03}. Therefore, Vela was weighted heavily in the search. Some of the outliers in Figure~\ref{plot_flux} (PSRs J$2021$$+$$4026$, J$1709$$-$$4429$, and J$1119$$-$$6127$) may be associated with poorly constrained distances. For an example, we used the kinematic distance of 1.5~kpc for the radio-quiet PSR J$2021$$+$$4026$ \citep{lrh+80}. The thermal emission component of the recent X-ray observations suggest a relatively large distance, $\sim 6$~kpc, at odds with a proposed association with supernova remnant G$78+2.1$ \citep{wrr+11}. These outliers are common for all of the methods that we follow in this paper (see Section~\ref{outer}, Figures~\ref{flux_tpc} and \ref{flux_og}).

\subsection{Period-dependent beaming factor}

There are numerous approaches for modeling the pulsar magnetosphere. The main approaches assume either the vacuum limit \citep{deu55} or force-free MHD limit \citep{spi06,tim06}. As suggested in \citet{lst+12}, the real pulsar magnetosphere is likely between these two limits. Considering particle corotation, the boundary of the magnetosphere is defined with respect to the light cylinder; $R_{\rm LC} = cP/2\pi$, where $c$ is the speed of light. Therefore, the emission geometry and pattern may depend on the pulsar period. Empirical fits to radio pulsars show that the half-opening angle $\rho_{\rm r}$ of the radio emission beam is a function of its period, $\rho_{\rm r} = 5\fdg8 P^{-0.5}$ \citep{ran93}. With this motivation, we model $f_\Omega = \lambda P ^{-\nu}$, including two additional parameters $\lambda$ and $\nu$. We vary the parameter $\nu$ from $-1$ to $3$, allowing $f_\Omega$ to have a maximum value of 1.5. Then we follow the same likelihood analysis with a grid search of all five parameters ($a$, $b$, $c$, $\lambda$, and $\nu$) to find marginalized PDFs. The best-fit parameters are $a = 0.1^{+1.2}_{-0.2}$, $b = 0.50\pm0.08$, $\log c = 33.3^{+0.1}_{-0.3}$, $\lambda = 2.2^{+0.2}_{-1.1}$, and $\nu = 1.8^{+0.2}_{-1.2}$. The PDFs of $a$, $\lambda$, and $\nu$ show that these cannot be constrained to single values, reflected in the larger errors (see Table~2; best-fit $\log(\chi^2_{\rm red}) = 5.1$). The non-zero $\nu$ may imply that $f_\Omega$ could have some period dependence with a more complex form dependent on additional parameters such as magnetic inclination and period derivative.

\subsection{Phase-shift dependent beaming factor}

\citet{gg01} and \citet{drh04} introduced a method to determine emission altitudes from core and conal emission components, as defined in \citet{ran83, ran83a, ran90, ran93}. For radio pulsars, the core component originates close to the surface and the conal component from a higher altitude. With this difference in emission altitude,  the leading conal component should be more widely separated from the main pulse than the trailing conal component due to aberration and retardation effects introduced by the pulsar rotation. Under this assumption, the emission height can be calculated from the phase shifts of these leading and trailing components. Similarly, for pulsars that are both radio and gamma-ray loud, it is possible that the phase shift between the radio and the gamma-ray pulse profile peaks can be explained with a similar effect,  leading to emission altitude estimates. Assuming that the high-energy emission is generated within the outer magnetosphere, the height of the  emission is related to the size of the instantaneous gamma-ray beam, and hence the beaming factor. Therefore, we calculate the phase shift of the leading and trailing components of the gamma-ray profile with respect to the radio pulse profile peak and explore whether the beaming factor depends on this quantity. Since the gamma-ray emission altitudes are larger than the radio emission altitudes, our assumption that the radio emission is generated on the neutron star surface is valid.

Assuming dipolar field lines, we can express the beam size as a power law of emission height, which is proportional to the  phase shift \citep[see][]{gg01}. Therefore, the beaming factor can be modeled as $f_\Omega=\lambda\Delta\phi^\nu / 4\pi$, where $\Delta\phi$ is the phase shift. We used only radio-loud GRPs that have multiple peaks in the gamma-ray pulse profile (19 out of 35 pulsar detections). We ignored the Crab pulsar (J$0534+2200$) due to nearly zero shift in phase between the peaks of the radio and gamma-ray pulse profiles, which implies that they arise from the same altitude. The best-fit parameters are $a = 2.0\pm0.2$, $b = 0.4\pm0.2$, $\log c = 32.0^{+0.4}_{-0.6}$, $\lambda = 5.2^{+14.6}_{-3.3}$, and $\nu = -0.4^{+0.3}_{-0.2}$. The parameter $\lambda$ is poorly constrained and its PDF shows that it has several equally significant solutions, resulting in a poor fit. Therefore, the phase-shift method is not applicable for gamma-ray and radio profiles, assuming a simple beam geometry. Including more parameters such as magnetic inclination with advanced emission geometries may lead to better fits.

\subsection{Outer magnetosphere model dependent beaming factor}
\label{outer}

Outer magnetosphere models can explain the broad profiles and observed emission bridges between pulse peaks in GRPs \citep{wr11}. The commonly used outer magnetosphere emission models are the OG model, in which the emission is generated within the gap region above the null charge surface ($\vec{B}\cdot\vec{\Omega}_{\rm rot} = 0$ where $\vec{\Omega}_{\rm rot}$ is the rotational angular momentum vector) extending toward the light cylinder, and the TPC model, in which the emission is generated within the gap region above the stellar surface extending toward the light cylinder.

The emission patterns produced by outer magnetosphere and PC models depend on the geometry of the pulsar, which can be determined from two main methods. For young pulsars with bright X-ray pulsar wind nebulae (PWNe), the viewing angle $\zeta$ can be constrained by fitting the Doppler-boosted PWN torus \citep{nr08}, where $\zeta$ is the angle between the line of sight and the rotation axis of the pulsar. However, since there are only a few pulsars with PWNe torii, the technique is limited. Radio polarization measurements, along with the rotating vector model \citep[RVM;][]{rc69a}, can also be used to constrain the magnetic inclination ($\alpha$) with respect to the rotation axis and the impact parameter ($\beta$), or the closest approach of the magnetic axis with respect to the line of sight. This model fits the S-shaped sweep of the polarization position angle of the linear polarization as a function of $\alpha$ and $\beta$.

Unfortunately, radio polarization has  not been measured for  all of the  pulsars in our study. \citet{ran93b} reported radio polarization measurements and  determined the geometrical angles $\alpha$ and $\beta$ for about 150 radio pulsars. \citet{mr11} further analyzed 50 pulsars that have asymmetric pulse profiles and polarization position angles and determined their geometry. We have calculated {\it Fermi} upper limits for 121 pulsars with known geometry from these studies. The geometry has been derived for only 14 of the LAT-detected pulsars with estimated distances; see Table~3. For these 14 pulsars, we calculate $f_\Omega$ for TPC and OG models using the analytical expressions derived in \citet{wrw+09} that models the beaming factor as a function of the pulsar geometry, $\alpha$ and $\zeta$($ = \beta+\alpha$), and the characteristic fractional gap width $w$. \citet{wrw+09} obtained $w$ from the assumption that $w\propto \eta=L_\gamma/\dot{E}$ with the luminosity form of $L_\gamma \propto \sqrt{\dot{E}}$. However, for $f_\Omega$ to be independent of the luminosity law, we assume that $w\approx0$, implying the gamma-ray emission is generated along the last closed field lines, which is consistent with original OG  \citep[e.g.][]{cr92,cr94} and TPC \citep{dr03} models. Therefore, we estimate $f_\Omega$ for all the 14 pulsars according to their geometries using expressions given in \citet{wrw+09} with $w\approx0$. For the rest of the 21 geometry-unknown detections, we use $f_\Omega = 1$.

\begin{table*}
\begin{center}
\caption{
Geometry estimated from radio polarization for {\it Fermi}-detected pulsars; the magnetic inclination $\alpha$, the impact parameter $\beta$, and half-opening angle of the radio beam $\rho_{\rm r}$ are listed.}
\begin{tabular}{lcccc}
\hline
\multicolumn{1}{c}{Pulsar} &
\multicolumn{1}{c}{$\alpha^{\dagger\dagger}$} &
\multicolumn{1}{c}{$\beta^{\dagger\dagger}$} &
\multicolumn{1}{c}{$\rho_{\rm r}$} &
\multicolumn{1}{c}{Ref} \\
\hline
J$0248+6021$ & $40\degr-80\degr$ & $+5\degr$ & $5\degr^\dagger$ & (1) \\
J$0631+1036$ & $90\degr$ & $-4\degr$ & $18\degr$ & (2) \\
J$0742-2822$ & $80\degr-110\degr$ & $-7\degr$ & $18\degr$ & (2) \\
J$0835-4510$ & $-137\degr$ & $6.5\degr$ & $6.5\degr^\dagger$ & (3) \\
J$1057-5226$ & $75\degr$ & $-6\degr$ & $6\degr^\dagger$ & (4), (5) \\
J$1119-6127$ & $20\degr-30\degr$ & $-30\degr - 0\degr$ & $14\degr$ & (6), (7) \\
J$1420-6048$ & $20\degr$ & $-0.5\degr$ & $15\degr$ & (2), (8) \\
J$1709-4429$ & $36\degr$ & $17\degr$ & $17\degr^\dagger$ & (5), (9) \\
J$1718-3825$ & $20\degr$ & $4\degr$ & $13\degr$ & (2) \\
J$2021+3651$ & $70\degr$ & $15\degr$ & $15\degr^\dagger$ & (10) \\
J$2030+3641$ & $20\degr - 90\degr$ & $20\degr - 80\degr$ & $5\degr^\dagger$ & (11) \\
J$2043+2740$ & $52\degr - 83\degr$ & $60\degr - 88\degr$ & $8\degr^\dagger$ & (12) \\
J$2229+6114$ & $55\degr$ & $-9\degr$ & $9\degr^\dagger$ & (13) \\
J$2240+5832$ & $108\degr$ & $123\degr$ & $15\degr^\dagger$ & (1) \\
\hline
\end{tabular}
\tablecomments{$^\dagger$ Half-opening angle $\rho_{\rm r}$ has not been constrained. Therefore, we assume $\rho_{\rm r} = |\beta|$ for any required calculation.\\
$^{\dagger\dagger}$ According to radio polarization, the two angles $\alpha$ and $\beta$ are associated with larger errors. \\
References: (1) \citet{tpc+11}; (2) \citet{waa+10}; (3) \citet{jhv+05}; (4) \citet{ww09}; (5) \citet{aaa+10b}; (6) \citet{wje+11}; (7) \citet{pkd+11}; (8) \citet{rrj+01}; (9) \citet{nr08}; (10) \citet{aaa+09b}; (11) \citet{ckr+12}; (12) \citet{naa12}; (13) \citet{aaa+09a}
}
\end{center}
\label{rvm}
\end{table*}

\begin{figure*}
\epsscale{1.70}
\plotone{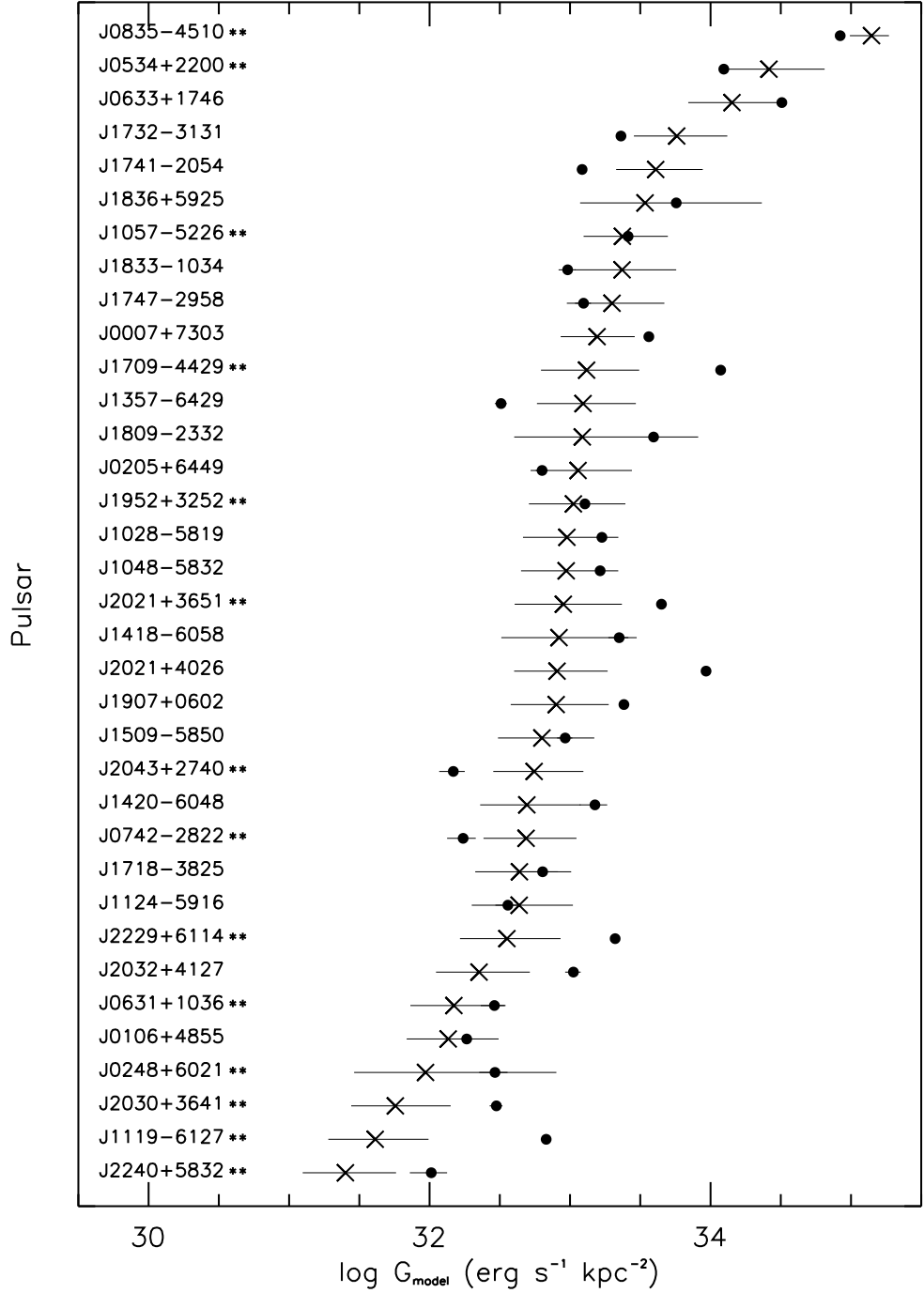}
\caption{
Same as Figure~\ref{plot_flux}, but for the TPC model. The best-fit luminosity model is $L_\gamma = 10^{32.81} P^{-1.45} \dot{P}^{0.41}$ (see Table~2). Note that the pulsars are marked with (**) have estimated geometry and included in the analysis. We used $f_\Omega=1$ for the pulsars that do not have estimated geometry.
\label{flux_tpc}}
\end{figure*}

For the TPC model, we use Equation (7) and (8) of \citet{wrw+09} to calculate $f_\Omega$ for the two cases $\zeta > \zeta_{\rm I}$ and $\zeta < \zeta_{\rm I}$ with $w=0.001$, where $\zeta_{\rm I} = (75+100w)-(60+1/w)(\alpha/90)^{2(1-w)}$ is Equation (6) of the same study. By searching over the three model parameters, we calculated the best-fit values to be $a = 1.45^{+0.04}_{-0.03}$, $b = 0.41\pm0.02$, and $\log c = 32.81^{+0.02}_{-0.03}$. The luminosity law is listed in Table~2 and model-estimated fluxes are shown in Figure~\ref{flux_tpc}.

\begin{figure*}
\epsscale{1.70}
\plotone{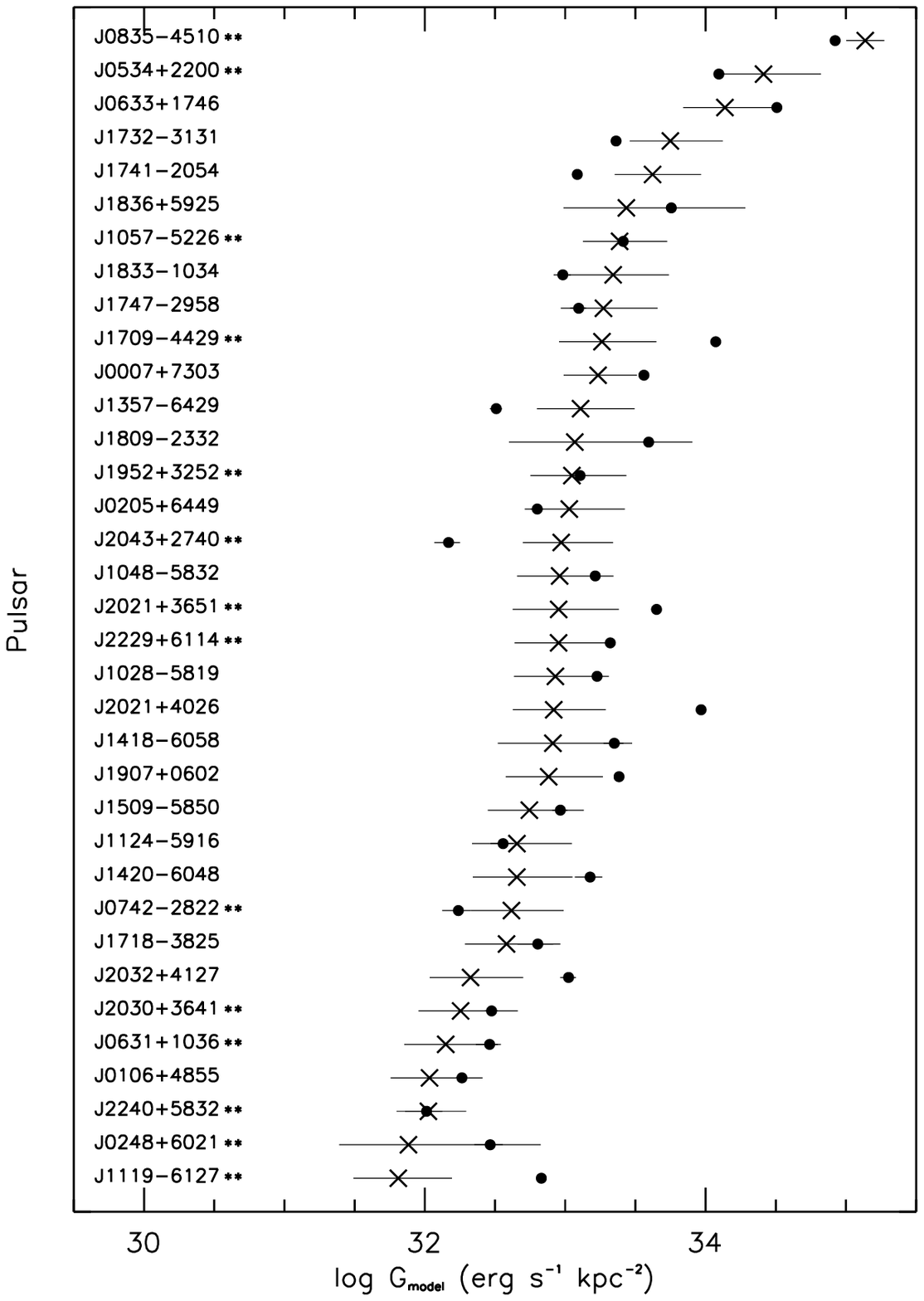}
\caption{
Same as Figure~\ref{plot_flux} and \ref{flux_tpc}, but for the OG model. The best-fit luminosity model is $L_\gamma = 10^{32.82} P^{-1.36} \dot{P}^{0.44}$ (see Table~2).
\label{flux_og}}
\end{figure*}

Then we constrain the luminosity law according to the OG model. Again, we use the derived results of \citet{wrw+09} with $w=0.001$. The corresponding analytic expressions for $f_\Omega$ for the OG model are given in Equations (9) and (10) in their study for the two cases of $\zeta > 60\degr$ and $\zeta < 60\degr$. The likelihood analysis then constrained the best-fit values to be $a = 1.36\pm0.03$, $b = 0.44\pm0.02$, and $\log c = 32.82^{+0.03}_{-0.02}$. The model-estimated and measured fluxes are shown in Figure~\ref{flux_og}.

\section{Determining the geometry of gamma-ray and radio beams}
\label{beam}

Gamma-ray and radio emission are likely generated in different regions of the magnetosphere for most pulsars, with gamma rays originating in the outer magnetosphere \citep{rom96a} and radio waves in the polar cap region \citep{rs75}. In general, the PC radio beam is assumed to be magnetic-pole-centered, while the gamma-ray emission geometry is complicated due to its outer magnetosphere origin. The sky maps (i.e. emission pattern in $\zeta$ vs $\phi_{\rm spin}$ space, where $\phi_{\rm spin}$ is the spin longitude) of the OG and TPC models \citep[see][]{bs10} show that the gamma-ray fan-like emission is centered on the spin equator (i.e. $\zeta \approx 90\degr$). In this section, we model the fraction of radio-detected pulsars with radio beams within their fan-like gamma-ray beams by using pulsar upper limits with our best-fit luminosity laws from different methods as described in the previous section.

We first calculate the number of observed severe non-detections and calculate the expected number given $\rho_\gamma$. The most likely explanation for a severe non-detection is that the gamma-ray beam does not intersect our line of sight due to radio and gamma-ray beam misalignment, with the radio beam outside of the gamma-ray beam. By applying the above condition to all upper-limit pulsars, we can estimate the number of {\it observed severe non-detections} ($N_{\rm \bar{\gamma}}$). The number of {\it expected severe non-detections} ($\hat{N}_{\rm \bar{\gamma}}$) can be estimated from $\rho_\gamma$. Then we use $\rho_\gamma$ to calculate the best-fit beaming solid angle $\Omega_\gamma$ of the gamma-ray emission.

We assume that the gamma-ray beam is centered around the spin equator and has a symmetric half-opening angle of $\rho_\gamma$, above and below the equator. However, the sky maps of OG and TPC models show that the gamma-ray emission pattern is strongly dependent on the magnetic inclination $\alpha$ \citep{bs10}. Therefore, the half-opening angle of the beam in our model is a function of $\alpha$. As shown in \citet{pgh+12}, the half-opening angle also depends on the age of the pulsar. Since we are interested in the average properties of the population, we do not include the age dependency on the half-opening angle in our simple model. The skew angle between the radio and gamma-ray beams is defined as $\theta_{\rm rg}$ (i.e., a polar angle measured from the spin equator). With the assumption that $\theta_{\rm rg}$ is uniformly distributed, we write the PDF as $f_{\rm \theta_{rg}}(\theta_{\rm rg}) \propto \sin\theta_{\rm rg}$ where $0\leq\theta_{\rm rg}<\rho_{\rm \gamma}$($\alpha$). Assuming that the radio beam size is small compared to that of the gamma-ray beam, we define the fraction of radio-detected pulsars that have gamma-ray beams that intersect our line of sight $f_{\rm r\gamma}$ as follows.

The beaming solid angle of the pulsar can be obtained by integrating over $\zeta$ and $\phi_{\rm spin}$ as
\begin{eqnarray}
\Omega_{\rm \gamma,i}(\alpha) = \int_0^{2\pi} d\phi_{\rm spin} \int_{(\pi/2 - \rho_{\rm \gamma}(\alpha))}^{(\pi/2 + \rho_{\rm \gamma}(\alpha))} \sin \zeta d\zeta. 
\end{eqnarray}

\noindent
By examining sky maps of \citet{bs10}, we find that, roughly, $\rho_{\rm \gamma,TPC}(\alpha) = 71\degr(\alpha/90\degr)^{0.7}$ and $\rho_{\rm \gamma,OG}(\alpha) = 75\degr(\alpha/90\degr)^{0.85}$ for TPC and OG models, respectively. Then the beaming fraction is given by $f_{\rm \Omega,i}(\alpha) \approx \Omega_{\rm \gamma,i}(\alpha)/4\pi$. We write $f_{\rm r\gamma}$ as a function of $\alpha$ as

\begin{eqnarray}
f_{\rm r\gamma}(\alpha) = \frac{\sum_{{\rm i=1}}^{N_{\rm r}}  \int_{0}^{\rho_{\rm \gamma}(\alpha)} f_{\rm \Omega,i}(\alpha) f_{\theta_{\rm rg}}(\theta_{\rm rg}) d\theta_{\rm rg}}{ \sum_{{\rm i=1}}^{N_{\rm r}}  \int_{0}^{\rho_{\rm \gamma}(\alpha)} f_{\theta_{\rm rg}}(\theta_{\rm rg}) d\theta_{\rm rg}}
\end{eqnarray}

\noindent
where $N_{\rm r}$ is the number of radio-detected pulsars in the sample. Then we estimate the number of expected severe non-detections for a given model from $\hat{N}_{\rm \bar{\gamma}} = (1-f_{\rm r \gamma}) N_{\rm r}$. For a given $\alpha$, we then calculate the likelihood function based on $\hat{N}_{\rm \bar{\gamma}}$. We define the likelihood fraction ${\cal L} = \exp (-0.5(N_{\rm \bar{\gamma}}-\hat{N}_{\rm \bar{\gamma}})^2 / N_{\rm \bar{\gamma}})$ when $\hat{N}_{\rm \bar{\gamma}} > 100$ and ${\cal L} = (\hat{N}_{\rm \bar{\gamma}})^{N_{\rm \bar{\gamma}}} \exp(-\hat{N}_{\rm \bar{\gamma}}) / N_{\rm \bar{\gamma}}!$ otherwise. We search $\alpha$ to find the best-fit value where the likelihood fraction is a maximum.

\begin{figure*}
\epsscale{1.80}
\plotone{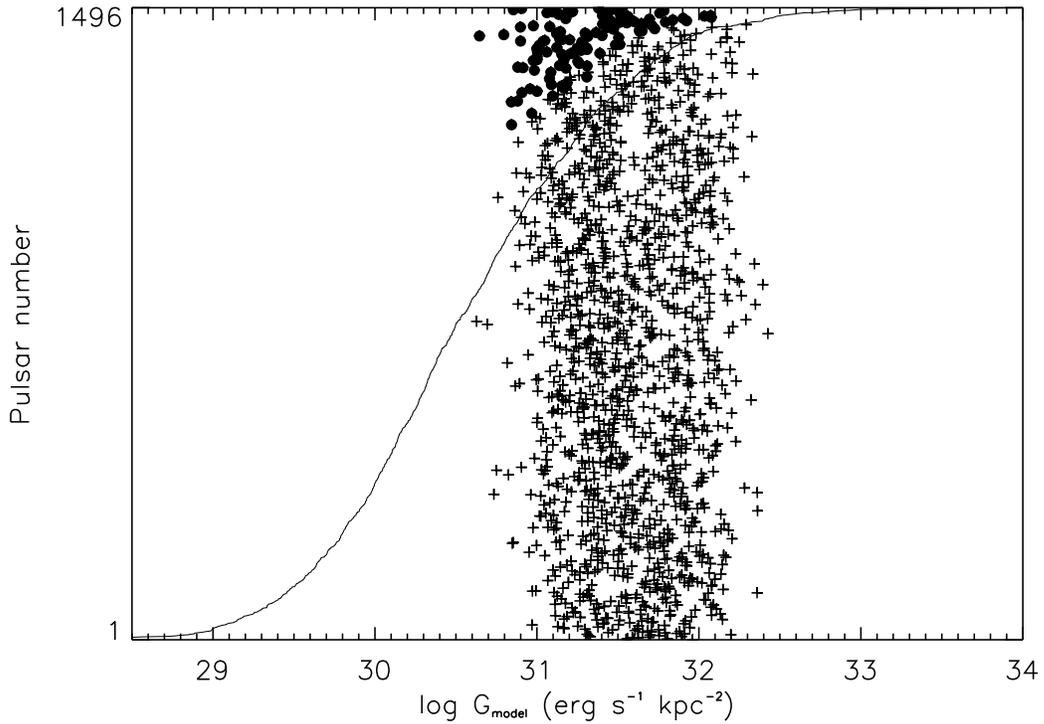}
\caption{
Model-estimated energy fluxes of 1496 non-recycled radio pulsars for the case of $f_\Omega=1$. The model-estimated fluxes are sorted according to decreasing order and connected by a thick line. Plus signs are the measured upper limit fluxes. The beaming analysis estimated 114 severe non-detections, which are located in the upper part of the figure and marked with filled circles. We define a severe non-detection as a measured upper limit that is 2$\sigma$ below the model prediction.
\label{upper}}
\end{figure*}

With the best-fit luminosity laws given in Section~\ref{results}, we determine $\alpha$, leading to $\rho_{\rm \gamma}$. For the model of $f_\Omega=1$, there are $114$ severe non-detections (i.e. where the expected flux is much greater than the measured upper limit flux). According to the explanation of a severe non-detection given above, about 92\% ($=[1-114/1496]\times100\%$, where 1496 is the total number of pulsars in the sample) of the radio-detected pulsars have their radio beams within the gamma-ray beams and are potentially detectable in gamma rays. Figure~\ref{upper} shows the measured upper limit fluxes of the known radio pulsars and their expected fluxes based on our best-fit luminosity law. Note that for the beaming analysis of this model, we use $\rho_{\rm \gamma}$, which is independent of $\alpha$, as the model parameter. However, we use the $\alpha$ dependence on $\rho_{\rm \gamma}$ in TPC and OG models below. By performing the above analysis, we constrained $\rho_{\rm \gamma} = 68(2)\degr$. Assuming a uniform beam, the corresponding $\rho_{\rm \gamma}$ implies that the average gamma-ray beaming solid angle is about $3.7(1)\pi$.

For outer magnetosphere models, we use the best-fit luminosity laws determined for TPC and OG models. The luminosity law of the TPC model predicts 162 severe non-detections. In other words, 11\% of the pulsars in the sample have their radio beams outside the gamma-ray beam. Note that for pulsars with estimated $\alpha$, we use that value directly to calculate the beaming solid angle of that pulsar. Fitting for the other pulsars, we find a best-fit $\alpha=77(4)\degr$, implying $\rho_{\rm \gamma, TPC}=64\degr$. The average beaming solid angle is then computed to be $\Omega_\gamma=3.6\pi$, assuming a uniform beam. Note that the errors for $\rho_{\rm \gamma}$ and $\Omega_\gamma$ are not quoted, because the derived $\rho_{\rm \gamma}(\alpha)$ expressions for TPC and OG models are approximates and associated with large errors. The best-fit luminosity law for the OG model predicts 117 severe non-detections (i.e. about 92\% of the upper-limit pulsars have their radio beams within the fan-like gamma-ray beams). The beaming analysis results in $\alpha=80(3)\degr$, so that $\rho_{\rm \gamma, OG}= 68\degr$. This gives an average total solid angle swept out by the gamma-ray beam of $3.7\pi$.

Note that \citet{wh11} predict an OG model death line for gamma-ray emission of $\log\dot{P} = -13.54 + 3.67\log P$. Using only the 561 upper limits for pulsars satisfying this criterion, we find 69 severe non-detections, implying that 88\% of the radio-detected pulsars have radio beams within the gamma-ray emission beams. The total solid angle swept out by the gamma-ray beam is $3.5\pi$ with a uniform beam.

\section{Properties and population analysis of GRPs with diffuse flux measurements}
\label{pop}

As in MC00, we construct a model pulsar population in the Galaxy and then assume that some fraction of the Galactic diffuse flux is due to unresolved pulsars. We use  our model-estimated flux in a given direction in the Galaxy and the measured diffuse flux in the same direction to place  constraints on braking index $n$, magnetic field $B_{12}$, and initial spin period $P_0$ of the Galactic pulsar population.

First we need a Galactic pulsar spatial distribution. MC00 used a simple model with a Gaussian disk, exponential halo, and molecular ring. In contrast, we use a more accurate spatial distribution derived in \citet{lfl+06} based on a sample of 1008 non-recycled pulsars from the Parkes multibeam survey. Then the pulsar number density $\rho(r,z)$ can be given as a function of radial distance from the Galactic center $r$ and the height from the Galactic plane $z$,

\begin{equation}
\label{spatial}
\rho(r,z) = \rho_{\rm 0} \left( \frac{r}{d_\odot} \right)^B \exp\left(\frac{-|z|}{E}\right) \exp \left[-C\left(\frac{r-d_\odot}{d_\odot}\right)\right]
\end{equation}

\noindent
where $E$, $B$, and $C$ are $0.18$~kpc, $1.9$, and $5.0$, respectively, and $\rho_{\rm 0}$ is a normalization constant. The distance to the Sun is taken to be $d_\odot=8.5$~kpc. According to \citet{agt+04} and \citet{lbdh93}, the Galactic pulsar birth rate  is likely between $1/125$~yr$^{-1}$ and $1/250$~yr$^{-1}$. We assume a conservative pulsar birth rate of $1/100$~yr$^{-1}$, implying the total number of pulsars in the Galaxy is $N_{\rm psr} = 10^8$. Thus, we find the normalization constant by integrating

\begin{equation}
N_{\rm psr} = \int_0^{r_0} dr \int_0^{2\pi} d\phi \int_{-z_0}^{z_0} dz r\rho(r,z)
\end{equation}

\noindent
where $r_0 = 30$~kpc and $z_0=10$~kpc, assuming that the Galaxy extends up to 30~kpc radially and 10~kpc above and below the plane. This integration returns $\rho_{\rm 0} = 4.05\times 10^5$~kpc$^{-3}$. Note that this particular distribution is cylindrically symmetric;  it does not depend on the azimuthal angle $\phi$, resulting in the same number of pulsars at a given radial distance for any $\phi$. Therefore, in order to compute the number of pulsars in a given Galactic direction, we convert the Galactic longitude ($l$) and latitude ($b$) to the above ($r$,$z$) coordinates using

\begin{eqnarray}
\label{convert}
&& x = D_{\rm i,\odot}\cos(b)\sin(l) \nonumber \\ 
&& y = d_{\odot} - D_{\rm i,\odot}\cos(b)\cos(l) \\
&& z = D_{\rm i,\odot}\sin(b) \nonumber \\ 
&& r = \sqrt{x^2 + y^2} \nonumber
\end{eqnarray}

\noindent
where $D_{\rm i,\odot}$ is the distance with respect to the Sun. We can then calculate the flux from pulsars in a volume element $dv$ at a distance $D_{\rm i,\odot}$ for a given Galactic direction as $dG = dv \rho(r,z) \langle L_\gamma | B_{12} \rangle / 4\pi f_\Omega D_{\rm i,\odot}^2$, where $\langle L_\gamma | B_{12} \rangle$ is the average luminosity of a pulsar for a given magnetic field. The expression for $\langle L_\gamma | B_{12} \rangle$ is given in Equation (\ref{ave_lum}). To simplify the analysis, according to our reference model discussed in Section~\ref{constant}, we fixed $f_\Omega$ to be unity for all pulsars in the Galaxy. The number of pulsars within a unit volume can be written as $dv\rho(r,z) = dD_{\rm i,\odot}A(D_{\rm i,\odot})\rho(r,z)$, where $A(D_{\rm i,\odot}) = \Omega_{\rm b} D_{\rm i,\odot}^2$, assuming a solid angle $\Omega_{\rm b}$ for the telescope beam centered on Galactic longitude $l$ and latitude $b$ with a uniform flux distribution across the beam. Then the model-estimated flux from pulsars along a Galactic direction ($l$,$b$) can be calculated as

\begin{eqnarray}
\label{diff_flux}      
G(l,b,\Omega_{\rm b})_{\rm model} &=& \int_{0}^{D_{\rm max}} dD_{\rm i,\odot} \frac{\langle L_\gamma | B_{12} \rangle}{\Omega D_{\rm i,\odot}^2} A(D_{\rm i,\odot}) \rho(r,z) \nonumber \\
&=& \langle L_\gamma | B_{12} \rangle \frac{\Omega_{\rm b}}{\Omega} \int_{0}^{D_{\rm max}} dD_{\rm i,\odot} \rho(r,z)
\end{eqnarray}

\noindent
where $D_{\rm max}$ is the distance of the furthest pulsar in a given Galactic direction ($l$,$b$) and the telescope beam solid angle $\Omega_{\rm b} = 2\pi (1-\cos\theta)$, where $\theta$ is the angular radius of the beam. We assumed a small $\theta$ of $1\degr$ in order to ensure an isotropic flux distribution across the beam. Therefore, with equations (\ref{ave_lum}), (\ref{spatial}), (\ref{convert}), and (\ref{diff_flux}), we can estimate the contribution from pulsars to the diffuse flux in a given Galactic direction.

We fit our model-estimated flux to the measured diffuse fluxes with a similar likelihood analysis as described in Section~\ref{lum}. In equation (\ref{ave_lum}), we use our best-fit $a$, $b$, and $c$ from the reference model to estimate the average luminosity of a pulsar for a given model $\Theta$. We evaluate the average luminosity according to the log-normal and flat distributions of the magnetic field and the initial spin period of the population, respectively. In order to do this, we follow

\begin{equation}
<L_\gamma|B_{12}> = \frac{\int dP_{\rm 0,i} \int dB_{\rm i} <L_\gamma|B_{\rm i}> \exp{(-(\log B_{\rm i} - B_{\rm 0})^2/2\sigma_{\rm i}^2)}}{\int dP_{\rm 0,i} \int dB_{\rm i} \exp{(-(\log B_{\rm i} - B_{\rm 0})^2/2\sigma_{\rm i}^2)}} 
\end{equation}

\noindent
where $B_{\rm i} \in [10^9,10^{15}]$~G and $B_{\rm 0} = \log(B_{12}\times 10^{12})$. We assumed $\sigma_{\rm i}=0.466$ by fitting a log-normal distribution to timing-derived surface magnetic fields of known non-recycled pulsars. We kept this value fixed because our analysis is not sensitive to $\sigma_{\rm i}$. However, we fit for the peak of the log-normal distribution. Recent studies estimate a wide range of initial spin periods $P_{\rm 0} \in [15$~ms$,150$~ms$]$ \citep[see, ][]{mgb+02,klh+03}, while \citet{wr11} preferred a short initial spin period $P_{\rm 0} \approx 50$~ms. Therefore, we assume that $P_{\rm 0}$ is bounded by $P_{\rm 0,min}$ and $P_{\rm 0,max}$, to be constrained in the analysis. In order to constrain $n$, $B_{12}$, $P_{\rm 0,min}$ and $P_{\rm 0,max}$, we follow a grid search for all four parameters while fitting the model-estimated diffuse flux to the measured diffuse flux with a step-like one-sided Gaussian function.

The contribution from pulsars to the Galactic diffuse flux is not well understood. Recent studies for high-latitude diffuse emission showed that the MSP population contributes a small fraction ($\sim 1\%$) of the Galactic diffuse emission \citep{aaa+12,gk13}. However, a similar study for Galactic plane emission has not been done. Therefore, to be conservative, we assume that a maximum of 10\% of the diffuse flux is due to pulsars. Furthermore, we discuss how this fraction affects our constrained values. For a given model $\Theta$, i.e. ($n$, $B_{12}$, $P_{\rm 0,min}$, $P_{\rm 0,max}$), the individual likelihood for the beam direction ($l$,$b$) is given as

\begin{equation}  
\label{L_diff_i}
{\cal L}_{\rm dif,i}(\Theta) = h_{\rm i}(G_{\rm i},\Theta),
\end{equation}

\noindent
where $h_{\rm i}(G_{\rm i},\Theta) = (2\pi\sigma_{\rm i}^2)^{-1/2} \exp (-(G(l,b,\Omega_{\rm b},\Theta)_{\rm mod,i} - \epsilon_{\rm d} G(l,b)_{\rm dif,i})/2\sigma_{\rm i} ^2)$ when $G(l,b,\Omega_{\rm b},\Theta)_{\rm mod} > \epsilon_{\rm d} G(l,b)_{\rm dif,i}$. If $G(l,b,\Omega_{\rm b},\Theta)_{\rm mod} < \epsilon_{\rm d} G(l,b)_{\rm dif,i}$ then the distribution $h_{\rm i}(G_{\rm i},\Theta) = (2\pi\sigma_{\rm i}^2)^{-1/2}$. Here, $G(l,b)_{\rm dif,i}$ and $\sigma_{\rm i}$ are the measured diffuse flux and its error, respectively. We define $\epsilon_{\rm d}$ as the fraction of the diffuse flux that is due to pulsars. Then we determine the total likelihood of the model $\Theta$ from ${\cal L}_{\rm dif}(\Theta) = \prod_{\rm i=1}^{N_{\rm beam}} {\cal L}_{\rm dif,i}(\Theta)$, where $N_{\rm beam}$ is the number of diffuse flux measurements, and then follow the same analysis to calculate the  PDFs of $n$, $B_{12}$, $P_{\rm 0,min}$, and $P_{\rm 0,max}$.

\begin{figure*}
\epsscale{2.0}
\plotone{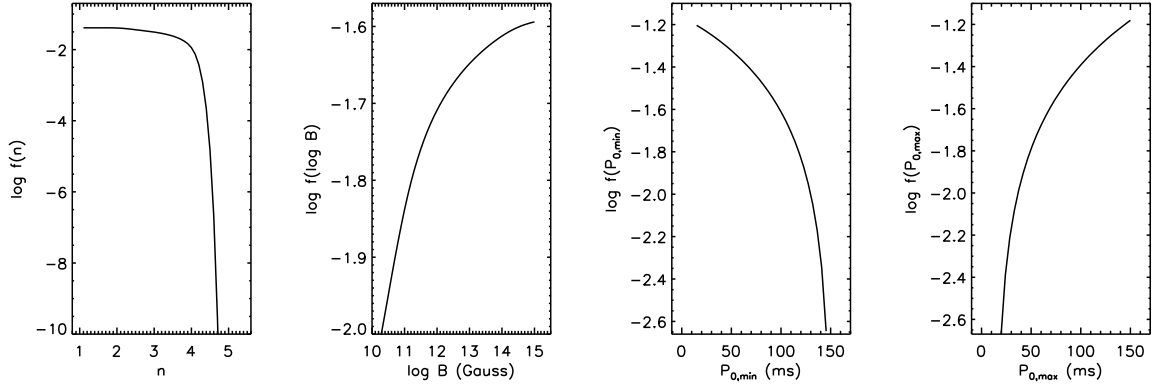}
\caption{
Marginalized PDFs for braking index $n$, surface magnetic field $B$, minimum limit of initial spin period $P_{\rm 0,min}$, and maximum limit of initial spin period $P_{\rm 0,max}$.
\label{diff}}
\end{figure*}

\begin{figure*}
\epsscale{1.60}
\plotone{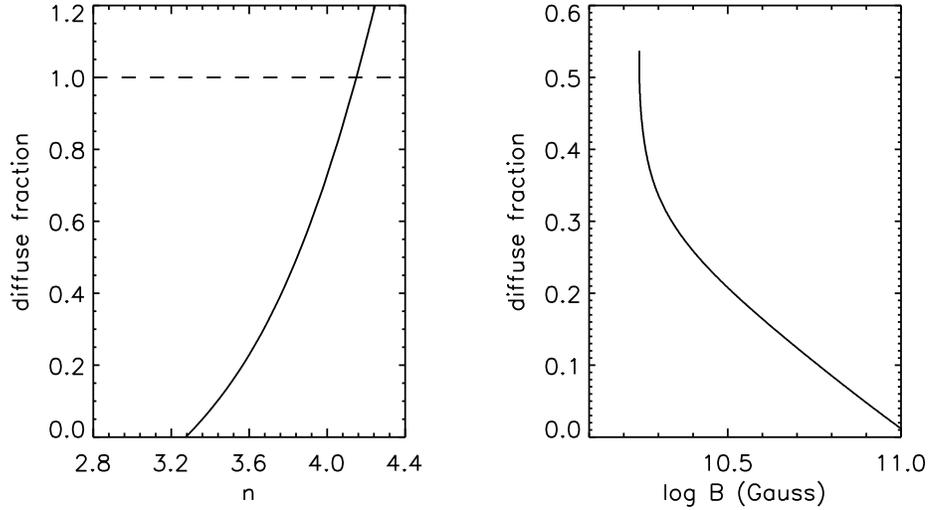}
\caption{
The fraction of Galactic diffuse flux due to unidentified pulsars as a function of braking index $n$ and surface magnetic field $B$. The curves of $n$ and $B$ are determined according to the upper and lower limits at 95\% confidence levels. In each plot, one parameter is varied while keeping the other parameter fixed, at a values of 2.5 for $n$ or $10^{12}$~G for $B$. The initial spin period $P_{\rm 0}$ is fixed at 100~ms throughout the fit. The dashed line shows where the fraction of diffuse flux due to pulsars reaches unity, showing that the range where $n>4.1$ is not allowed.
\label{diff_frac}}
\end{figure*}

With the assumptions that $f_\Omega=1$ and $\epsilon_{\rm d} = 0.1$, and assuming the best-fit parameters of $a$, $b$, and $c$ as determined in Section~\ref{constant}, we search the entire parameter space of $n$, $B_{12}$, $P_{\rm 0,min}$ and $P_{\rm 0,max}$ and calculate their PDFs, shown in Figure~\ref{diff}. We place a 95\%-confidence upper-limit on $n$ and $P_{\rm 0,min}$ of $3.8$ and $120$~ms, respectively, and a 95\%-confidence lower-limit on $B$ and $P_{\rm 0,max}$ of $3.2\times 10^{10}$~G and $46$~ms, respectively. Changing $\epsilon_{\rm d}$ alters these limits slightly. For $\epsilon_{\rm d} = 0.01$, we find a 95\%-confidence upper-limit on $n$ of 3.7 and a 95\%-confidence lower-limit on $B$ to be $2.2 \times 10^{10}$~G. Clearly, these limits are not constraining the pulsar population. In Figure~\ref{diff_frac}, we show how the fraction of diffuse flux due to unidentified pulsars varies with $n$ and $B$. Note that the curves of $n$ and $B$ are determined according to the upper and lower limits at 95\% confidence levels, respectively. We fixed $P_{\rm 0}$ at 100~ms to be consistent with the results of \citet{pt12}. We vary only one parameter at a time while keeping the other parameters at their typical values of $n=2.5$ and $B = 10^{12}$~G. 
This clearly shows that the diffuse fraction due to pulsars is not constraining. 
Furthermore, we found the diffuse fraction due to pulsars in the direction towards the Galactic center is about 0.6\% with typical parameters $n=2.5$, $B = 10^{12}$~G, and $P_{\rm 0}=100$~ms.

\begin{figure*}
\epsscale{1.65}
\plotone{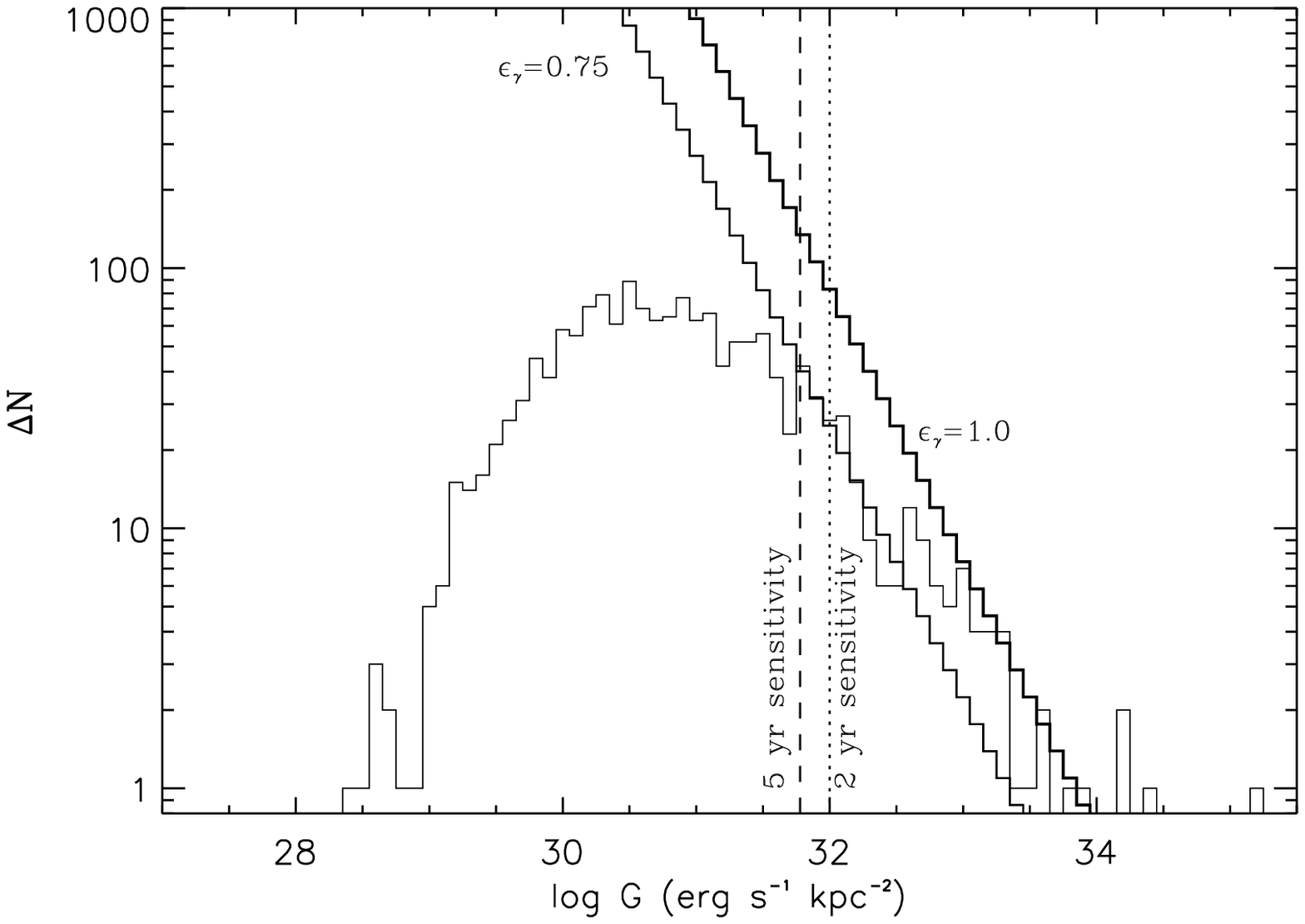}
\caption{
Predicted {\it Fermi} LAT energy fluxes of the GRP population model. The thin solid line is the model-predicted fluxes for identified non-recycled pulsars according to their period and period derivative with our best-fit luminosity law for $f_\Omega=1$. The two thick solid curves are the predicted fluxes for the model population of GRPs for two different maximum efficiencies $\epsilon_\gamma$. The dotted line is the LAT 2 year sensitivity for a point source in the Galactic-plane \citep{naa+12}. The dashed line is the expected LAT 5 year point source sensitivity.
\label{GRP}}
\end{figure*}

Using Equation~(\ref{num_pul}), we can estimate the flux distribution of GRPs in the Galaxy with the best-fit luminosity law and $n=2.5$, $B=10^{12}$~G, and $P_{\rm 0} = 100$~ms (see Figure~\ref{GRP}). We derive two different model population predicted fluxes for $\epsilon_\gamma = 0.75$ and $1$. We then predict the number of detectable pulsars using the LAT according to its sensitivity. These are discussed in the next section.

\begin{figure*}
\epsscale{1.60}
\plotone{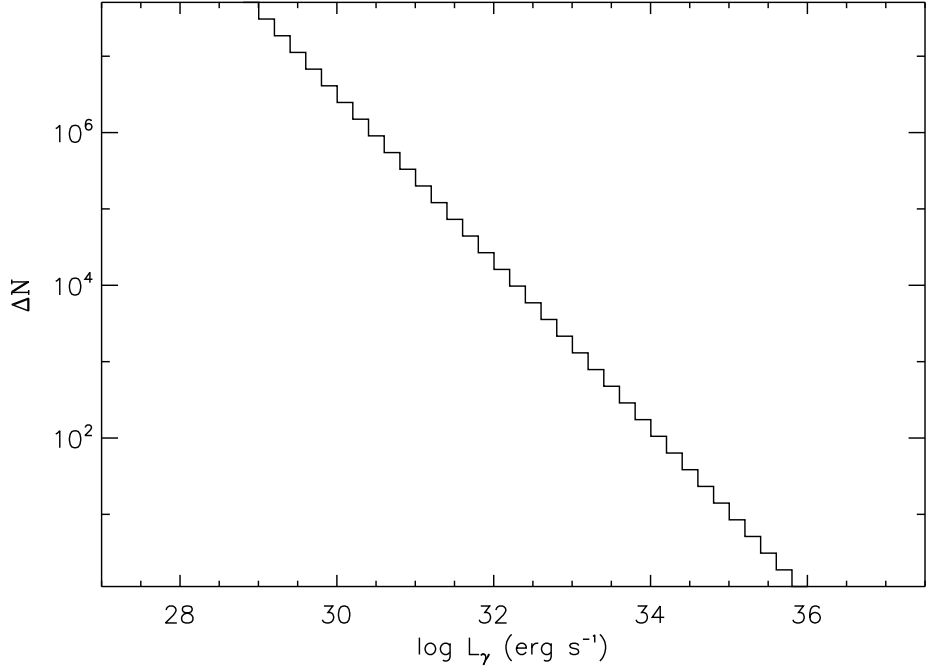}
\caption{
The luminosity distribution of GRPs in the Galaxy. The best-fit luminosity function is constrained to be $\log N(L_\gamma) = -1.09 \log L_\gamma + 39.3$. Note that we assumed all the pulsars have a beaming factor of $f_\Omega=1$.
\label{lum_pop}}
\end{figure*}

Furthermore, we model the luminosity distribution of GRPs in the Galaxy. Assuming that all the pulsars have $f_\Omega=1$, we use the corresponding best-fit parameters $a=1.43$, $b=0.40$, and $\log(c)=32.83$ (see Table~2), with $n=2.5$, $B=10^{12}$~G, and assume a low initial spin period $P_{\rm 0} = 15$~ms. Then we write the differential number of pulsars versus luminosity $dN_{\rm psr}/dL_\gamma = N_{\rm psr} f_{\rm L_\gamma}(L_\gamma|B_{12})$, where $f_{\rm L_\gamma}(L_\gamma|B_{12})$ is given in Equation~(\ref{pdf_f_L}). With a similar form of expression as given in Equation~(\ref{dn}), but for the luminosity, we estimate the number of pulsars for a given luminosity. Figure~\ref{lum_pop} shows the Galactic luminosity distribution. The luminosity function is then fit with a power law as

\begin{equation}
N(L_\gamma) = \left\{ \begin{array}{cl}
k L_\gamma^{-m} &\mbox{for $L_{\rm \gamma,min} < L_\gamma < L_{\rm \gamma,max}$} \\
0 &\mbox{for otherwise}
\end{array} \right.
\end{equation}

\noindent
where $m = 1.09$ and $\log k = 39.3$. The minimum and maximum luminosities $L_{\rm \gamma,min} = 8\times 10^{28}$~erg s$^{-1}$ and $L_{\rm \gamma,max} = 1.6\times 10^{36}$~erg s$^{-1}$ are determined from the oldest (from Equation~\ref{period_int}) and the youngest pulsar in the Galaxy based on their spin periods, respectively.

\section{Discussion}
\label{dis}

The best-fit luminosity laws show a significant improvement in uncertainties of the parameters compared to the results of previous works, MC00 and MC03, due to the larger number of GRPs in our sample. 
We find the constant $f_\Omega$ model provides a best-fit luminosity law that closely follows the canonical form $L_\gamma \propto \sqrt{\dot{E}}$. 
In order to include the gamma-ray emission models in the analysis, we then constrained the luminosity law based on the geometry-dependent OG and TPC models. 
However, these provided similar results to the constant $f_\Omega$ model because they are consistent with wide gamma-ray beams, so that $f_\Omega$ is almost equal to unity.
However, the OG model ($L_\gamma \propto P^{-a} \dot{P}_{\rm 15}^{b}$ where $a=1.36\pm0.03$ and $b=0.44\pm0.02$) provides the lowest chi-squared value among these three models and therefore, we claim it gives the best-fit to the data in general.
This is consistent with the result of \citet{wr11} that the OG model is strongly preferred in order to explain the observed spin and pulse properties of GRPs.  The data are not well fit through a PC model (see Table~2). We use the best-fit luminosity law for $f_\Omega=1$ model as the reference model for the upper-limit analysis and population model. The large chi-squared values associated with our fits, however, imply that the analyses are dominated by distance and beaming uncertainties. Some of the outliers in Figure~\ref{plot_flux}, \ref{flux_tpc}, and \ref{flux_og} may be associated with poor distance estimates. Parallax measurements for additional pulsars would dramatically improve our analysis.

Using flux upper limits, we determined the number of radio-detected pulsars that cannot be observed in gamma rays due to their large beam misalignment. The OG model predicts 117 such severe non-detections, or 8\% of the population. Although our simple beam model does not perfectly represent gamma-ray beam geometries, it constrains the relationship of radio and gamma-ray beams and offers insights into the numbers of radio-loud and radio-quiet gamma-ray pulsars. Furthermore, the beam model predicts large beaming solid angles, implying that the pulsar gamma-ray emission covers almost the entire sky, consistent with realistic outer magnetosphere emission models.

Using LAT diffuse fluxes and our population model, we can constrain some properties of the GRP population such as $n$, $B_{12}$, and $P_0$. A 2$\sigma$ upper limit on the average $n$ and a 2$\sigma$ lower limit on the average $B_{12}$ of the population are $3.8$ and $3.2 \times 10^{10}$~G, respectively. While these are not yet physically constraining, they should become so with time as more point sources are discovered and more accurate diffuse background models are developed.

According to our reference luminosity law along with the periods and period derivatives of known pulsars, the mean of the gamma-ray flux distribution is about $2 \times 10^{32}$~erg s$^{-1}$ kpc$^{-2}$ (see Figure~\ref{GRP}). We then calculate the number of LAT-detectable pulsars based on our Galactic model with a maximum gamma-ray efficiency of $\epsilon_\gamma =1$. Given the LAT sensitivity for a point source in the Galactic plane ($|l|<1\degr$) of $1\times 10^{32}$~erg s$^{-1}$ kpc$^{-2}$ from the 2FGL based on the first 24 months of data \citep{naa+12}, it is capable of detecting about 380 non-recycled pulsars as point sources, including 150 currently identified pulsars. 
According to the 2PC, pulsations from 77 non-recycled pulsars have been reported.
With the expected 5-year point source sensitivity of $6\times 10^{31}$~erg s$^{-1}$ kpc$^{-2}$, scaled from the 2-year sensitivity (i.e., $1\times 10^{32}$~erg s$^{-1}$ kpc$^{-2}$ times $\sqrt{2/5}$),  it is capable of detecting emission from about 620 pulsars, including about 220 currently identified pulsars.

These prediction numbers are larger than the number of detections reported in the 2PC. The most likely explanation for this is that the efficiency of detection falls off at lower sensitivity levels. This can be clearly shown by looking at the flux distribution of the 2PC detections, which peaks at a flux roughly five times higher than the quoted sensitivity limit. In addition, note that our model assumptions will impact the predictions. For instance, a smaller beaming fraction would imply smaller numbers of detectable GRPs. Also the assumed Galactic spatial distribution has large errors. Therefore, more detailed analysis can be done in the future including better distance estimates, a more accurate proper beaming model, and more accurate distributions of periods, inclination angles, age, etc.

\bigskip

\acknowledgments

BBPP and MAM are supported through the Research Corporation. MAM gratefully acknowledges support from Oxford Astrophysics while on sabbatical leave. AKH acknowledges support from the Fermi GI and NASA Astrophysics Theory Programs.
Some of the results in this paper have been derived using the HEALPix \citep{ghb+05} package.

The \textit{Fermi} LAT Collaboration acknowledges generous ongoing support from a number of agencies and institutes that have supported both the development and the operation of the LAT as well as scientific data analysis. These include the National Aeronautics and Space Administration and the Department of Energy in the United States, the Commissariat \`a l'Energie Atomique and the Centre National de la Recherche Scientifique / Institut National de Physique Nucl\'eaire et de Physique des Particules in France, the Agenzia Spaziale Italiana and the Istituto Nazionale di Fisica Nucleare in Italy, the Ministry of Education, Culture, Sports, Science and Technology (MEXT), High Energy Accelerator Research Organization (KEK) and Japan Aerospace Exploration Agency (JAXA) in Japan, and the K.~A.~Wallenberg Foundation, the Swedish Research Council and the Swedish National Space Board in Sweden. Additional support for science analysis during the operations phase is gratefully acknowledged from the Istituto Nazionale di Astrofisica in Italy and the Centre National d'\'Etudes Spatiales in France.

\appendix
\section{Appendix}
For a given luminosity law with an assumed spin-down law $\dot{\Omega} \propto \Omega^n$, where $\Omega$ is the spin frequency and $n$ is the braking index, the average luminosity of a pulsar for a given magnetic field strength can be written as 

%\begin{equation}
%\label{ave_lum}
%\langle L_\gamma | B_{12} \rangle  = \frac{2 c B_{12}^{b} P_{\rm 0}^{(2-a)}}{P_{\rm g}^2 (a -2)} \left[ 1- \left( 1+ \frac{T_{\rm g}}{\tau_{\rm 0}} \right)^{(2-a)/(n-1)} \right]
%\end{equation} 

\begin{equation}
\label{ave_lum}
\langle L_\gamma | B_{12} \rangle  = \frac{2 c B_{12}^{2b} P_{\rm 0}^{(2-a-b)}}{P_{\rm g}^2 (a +b -2)} \left[ 1- \left( 1+ \frac{T_{\rm g}}{\tau_{\rm 0}} \right)^{(2-a-b)/(n-1)} \right]
\end{equation}

\noindent
for $a \neq 2$, $n \neq 1$, and

%\begin{equation}
%\langle L_\gamma | B_{12} \rangle  \approx \frac{10^{15} c B_{12}^{b-2} P_{\rm 0}^{2-a}}{T_{\rm g} (a-2)}
%\end{equation}

\begin{equation}
\langle L_\gamma | B_{12} \rangle  \approx \frac{10^{15} c B_{12}^{2(b-1)} P_{\rm 0}^{2-a-b}}{T_{\rm g} (a+b-2)}
\end{equation}

\noindent
for $a > 2$ and $n=3$ (refer to Equations (7) and (8) in MC00). In these equations, the initial spin-down time $\tau_{\rm 0}$ and the period of the oldest pulsar in the Galaxy $P_{\rm g}$ can be given by

\begin{eqnarray}
\label{period_int}
&& \tau_{\rm 0} = \frac{10^{15} P_{\rm 0}^2}{B_{12}^2 (n-1)} \\
&& P_{\rm g} = P_{\rm 0} \left( 1 + \frac{T_{\rm g}}{\tau_{\rm 0}} \right)^{1/(n-1)} \nonumber
\end{eqnarray}

\noindent
where $P_{\rm 0}$ and $T_{\rm g}$ are the initial spin period of the pulsar and the age of the Galaxy ($T_{\rm g} = 10^{10}$~yr), respectively.

For a Galactic pulsar population, the number of pulsars for a given flux $G$ can be written as follows

\begin{equation}    
\label{num_pul}
\Delta N_{\rm psr}(G) = 2G\sinh(1.15\Delta \log G) \frac{dN_{\rm psr}}{dG}
\end{equation}

\noindent
where $\Delta \log G$ is the logarithmic bin size of the flux. The differential number of pulsars with respect to flux can be given as

\begin{equation}
\label{dn}
\frac{dN_{\rm psr}}{dG} = \frac{\Omega}{4\pi} N_{\rm psr} \int dL_\gamma f_{\rm L_\gamma}(L_\gamma|B_{12})f_{\rm G}(G|L_\gamma)
\end{equation}

\noindent
where the PDFs $f_{\rm L_\gamma}(L_\gamma|B_{12})$ and $f_{\rm G}(G|L_\gamma)$ are given by

%\begin{eqnarray}
%\label{pdf_f_L}
%f_{\rm L_\gamma}(L_\gamma|B_{12}) = \frac{2 c^{2/a}}{a P_{\rm g}^2}B_{12}^{2 b/a} L_\gamma^{-(1+2/a)} \\
%f_{\rm G}(G|L_\gamma) = \frac{1}{2}\left( \frac{L_\gamma}{\Omega} \right)^{1/2} f_{\rm D}(D_{\rm i,\odot}) G^{-3/2}.
%\end{eqnarray}

\begin{eqnarray}
\label{pdf_f_L}
f_{\rm L_\gamma}(L_\gamma|B_{12}) = \frac{2 c^{2/(a+b)}}{(a+b) P_{\rm g}^2}B_{12}^{4 b/(a+b)} L_\gamma^{-(1+2/(a+b))} \\
f_{\rm G}(G|L_\gamma) = \frac{1}{2}\left( \frac{L_\gamma}{\Omega} \right)^{1/2} f_{\rm D}(D_{\rm i,\odot}) G^{-3/2}.
\end{eqnarray}

\noindent
Here, $f_{\rm D}(D_{\rm i,\odot})$ is the PDF of the spatial distribution of the Galactic pulsar population as a function of the distance from the Sun $D_{\rm i,\odot}$. Assuming  cylindrical symmetry around the Galactic center, we can write the PDF

\begin{equation}
f_{\rm D}(D_{\rm i,\odot}) = A\frac{D_{\rm i,\odot}}{2\pi d_\odot} \int\int dr dz \rho(r,z)\frac{1}{r}\left[ 1-\left( \frac{r^2+z^2+d_\odot^2-D_{\rm i,\odot}^2}{2rd_\odot}\right)^2 \right]^{-1/2},
\end{equation}

\noindent
where $|r^2+z^2+d_\odot^2-D_{\rm i,\odot}^2| \leq 2rd_\odot$ for all $r$ and $z$. The normalization constant $A$ can be evaluated with the condition that $\int dD_{\rm i,\odot} f_{\rm D}(D_{\rm i,\odot})=1$ and the pulsar number density $\rho(r,z)$ is given in Equation (\ref{spatial}). All the derivations of these equations are given in MC00.

\bibliographystyle{apj.bst}
\bibliography{psrrefs,modrefs,journals}

\end{document}